\documentclass[twocolumn]{aastex631}

\newcommand\chandra{{\it Chandra} }
\newcommand{\ha}{$\rm H\alpha$ }

\usepackage{bm}
\usepackage{graphicx}
\shorttitle{M31 nucleus}
\shortauthors{Su et al.}
\graphicspath{{./}{figures/}}

\defcitealias{2003ApJ...599..237P}{PT03}

\begin{document}

\title{Wind-fed Supermassive Black Hole Accretion by the Nuclear Star Cluster: the Case of M31*}

\author[0000-0002-6738-3259]{Zhao Su}
\affiliation{School of Astronomy and Space Science, Nanjing University, Nanjing 210046, China}
\affiliation{Key Laboratory of Modern Astronomy and Astrophysics (Nanjing University), Ministry of Education, Nanjing 210046, China}
\email{suzhao@smail.nju.edu.cn}

\author[0000-0003-0355-6437]{Zhiyuan Li}
\affiliation{School of Astronomy and Space Science, Nanjing University, Nanjing 210046, China}
\affiliation{Key Laboratory of Modern Astronomy and Astrophysics (Nanjing University), Ministry of Education, Nanjing 210046, China}
\affiliation{Institute of Science and Technology for Deep Space Exploration, Suzhou Campus, Nanjing University, Suzhou 215163, China}
\email{lizy@nju.edu.cn}
\author[0000-0002-7172-6306]{Zongnan Li}
\email{zongnan.li@astro.nao.ac.jp}
\affiliation{National Astronomical Observatory of Japan, 2-21-1 Osawa, Mitaka, Tokyo, 181-8588, Japan}
\affiliation{East Asian Core Observatories Association (EACOA) Fellow}


\begin{abstract}
The central supermassive black hole (SMBH) of the Andromeda galaxy, known as M31*, exhibits dim electromagnetic emission and is inferred to have an extremely low accretion rate for its remarkable mass ($\sim10^8~\rm~M_\odot$).
In this work, we use 
three-dimensional hydrodynamical simulations to explore a previously untested scenario, 
in which M31* is fed by the collective stellar mass-loss from its surrounding nuclear star cluster, manifested as a famous eccentric disk of predominantly old stellar populations.
The stellar mass-loss is assumed to be dominated by the slow and cold winds from $100$ asymptotic giant-branch stars, which follow well-constrained Keplerian orbits around M31* and together provide a mass injection rate of 
$\sim4\times10^{-5}\rm~M_\odot~yr^{-1}$. The simulations achieve a quasi-steady state on a Myr timescale, 
at which point a quasi-Keplerian, cool ($T\sim10^3-10^4~\rm K$) gas disk extending several parsecs is established. This disk is continuously supplied by the stellar winds and itself feeds the central SMBH. 
At the end of the simulations at 2 Myr, 
an accretion rate of $\sim2\times10^{-5}\rm~M_\odot~yr^{-1}$ is found but could vary by a factor of few depending on whether the subdominant gravity of the NSC or a moderate global inflow is included. 
The predicted X-ray luminosity of $\sim10^{36}~\rm erg~s^{-1}$, dominated by the hot ($T\sim10^7-10^8~\rm K$) plasma within 0.2 parsec of the SMBH, is well consistent with {\it Chandra} observations.
We conclude that the feeding mechanism of M31* is successfully identified, which has important implications for the working of dormant SMBHs prevalent in the local universe. 
\end{abstract}



\section{Introduction}\label{sec:intro}
Supermassive black holes (SMBHs) stay at a low state of activity throughout most of their lifetimes, manifested as low luminosity active galactic nuclei \citep[LLAGN;][]{2008ARA&A..46..475H}.
Although their accretion power and energy output are much weaker than those of actively accreting SMBHs, LLAGNs can still significantly impact the evolution of their host galaxies through the so-called ``hot mode" feedback \citep[e.g.,][]{2017MNRAS.465.3291W,2018MNRAS.479.4056W,2019ApJ...885...16Y}.
However, the mechanisms by which the LLAGNs are fed and regulated remain poorly understood.

Stellar winds from nuclear star clusters (NSCs) could be one of the most immediate and direct supplies for the accretion onto SMBHs.
NSCs are extremely dense and compact stellar systems residing in the center of most galaxies, which commonly co-exist and likely co-evolve with the central SMBHs \citep[see review by][]{2020A&ARv..28....4N}.
This wind-feeding scenario is well established for Sgr A*, the SMBH of our own Galaxy, thanks to extensive theoretical studies and numerical simulations based on well-constrained stellar properties of the Milky Way NSC \citep[e.g.,][]{2004ApJ...613..322Q,2006MNRAS.366..358C,2018MNRAS.478.3544R,2020ApJ...888L...2C}.
Specifically, Sgr A* is found to be fueled by the winds of several tens of Wolf-Rayet stars embedded within the central parsec, which are the most massive members of the NSC \citep[e.g.,][]{2018MNRAS.478.3544R,2020MNRAS.492.3272R,2020ApJ...896L...6R}.
The fast winds ($\sim 1000-3000~\rm km~s^{-1}$) having mass-loss rates of $\sim10^{-5}~\rm M_\odot~\rm yr^{-1}$ from these Wolf-Rayet stars sculpt the hot gas flow in the inner parsec by wind-wind collisions and collectively drive a nuclear outflow.
Yet a minor fraction ($\lesssim10^{-3}$) of the stellar winds is trapped by Sgr~A*, forming a {\it Bondi}-like accretion, producing thermal X-ray emission consistent with {\it Chandra} observations of Sgr A* in its quiescent state \citep{2013Sci...341..981W, 2017MNRAS.464.4958R, 2018MNRAS.478.3544R, 2024ApJ...974...99B}.

Although the wind-feeding scenario well reproduces the observations of Sgr A*, the case of Sgr A* may be atypical for the young, massive stars (3--8 Myr) with extreme mass-loss in the immediate vicinity of the SMBH \citep{2006ApJ...643.1011P, 2015A&A...584A...2F}.
By contrast, many NSCs are observed to host a predominantly old population ($\rm >1~\rm Gyr$), and a minor component of intermediate-age stars ($\sim100-300~\rm Myr$) \citep{2006ApJ...649..692W, 2015AJ....149..170C, 2018MNRAS.480.1973K}.
Therefore, SMBHs embedded in NSCs deficient of young stars ($\lesssim10~\rm Myr$) and recent star formation may constitute a distinct case for the wind-feeding scenario, in which accretion is primarily fed by mass loss from the dominant old stellar population.

Little attention has been paid to such a scenario, however.
\citet{2009ApJ...699..626H} estimated the stellar mass loss from old, evolved stars in typical galactic nuclei and found that it is sufficient to account for the observed low accretion rate (represented by $L_{\rm bol}/L_{\rm Edd}\lesssim10^{-3}$, where $L_{\rm bol}$ and $L_{\rm Edd}$ are the bolometric luminosity and Eddington luminosity, respectively) in most nearby SMBHs. 
\citet{2014ApJ...782..103S} used one-dimensional numerical simulations to model the accretion flow of the LLAGN in NGC\,3115 fed by a 5-Gyr-old nuclear stellar population, which was found to be consistent with {\it Chandra} X-ray observations. This was supported by \citet{2020MNRAS.492..444Y} using two-dimensional simulations.

The SMBH of the Andromeda galaxy (M31), known as M31*, is one of the nearest and most underluminous SMBHs known,
making it an important object to investigate the radiatively inefficient accretion flow and symbiotic outflows of SMBHs at low accretion rates, alongside Sgr A* \citep{2014ARA&A..52..529Y}.
M31* has a dynamical mass of $\sim1.4\times10^{8}~\rm M_\odot$ \citep{2005ApJ...631..280B}, corresponding to an Eddington luminosity of $\sim10^{46}~\rm erg~s^{-1}$.
First detected in the radio band decades ago \citep{1992ApJ...390L...9C,1993ApJ...417L..61C}, M31* remains a compact source over the centimeter wavelengths 
with a steep spectrum 
hinting at optically thin synchrotron emission from a putative jet or wind powered by a hot accretion flow \citep{2017ApJ...845..140Y,2023ApJ...953...12P}.
In the X-ray band, sensitive \chandra observations identified an X-ray counterpart to M31* and determined the X-ray luminosity to be $\sim10^{36}~\rm erg~s^{-1}$ \citep{2005ApJ...632.1042G,2009MNRAS.397..148L}, indicating a bolometric luminosity of only $\sim 10^{-9}~L_{\rm Edd}$.
Notably, M31* exhibited a flare in 2006 that was $\gtrsim100$ times brighter than its quiescent state, which was followed by a more active phase with elevated and variable X-ray emission \citep{2011ApJ...728L..10L}.
\citet{2025ApJ...981...50D} found that this elevated phase persisted through 2024 and also revealed a moderate flare in 2013.
The strong X-ray variability of M31* is reminiscent of the X-ray flares of Sgr A* whose physical origin remains elusive.

Due to its proximity, the NSC of M31 is one of the few NSCs for which the stellar population and dynamics can be studied in detail. 
The most remarkable feature of this NSC, which occupies the inner few parsecs of M31, is that it is dominated by the so-called double-nucleus, designated by P1 and P2 \citep{1993AJ....106.1436L,1998AJ....116.2263L}. 
The double-nucleus morphology, as well as the stellar velocity field, is well explained by an eccentric disk of old, metal-rich stars on close Keplerian orbits around M31* \citep{1995AJ....110..628T, 2003ApJ...599..237P}, which has an inferred total mass of $\sim2\times10^{7}~\rm M_\odot$ \citep{2010A&A...509A..61S,2018ApJ...854..121L}.
The origin and long-term stability of this eccentric stellar disk, however, remains elusive.
In addition to the old population, a compact group of bluer stars,  designated by P3, is found in the very vicinity of the SMBH, the collective spectrum of which is consistent with A-type stars \citep{2005ApJ...631..280B,2012ApJ...745..121L}. 

The presence of P3 suggests a burst of star formation 100--200 Myr ago in the M31 nucleus, with an initial mass of a few thousand solar masses. 
However, the potential fuel for this starburst is not immediately clear, as the central 10--100 parsec region of M31 is known to have little amount of cold gas \citep[][]{2013A&A...549A..27M, 2017A&A...607L...7M, 2019MNRAS.484..964L}.
\citet{2007ApJ...668..236C} proposed that the stellar mass loss from the old eccentric disk, gravitationally bounded to M31*, would accumulate into a gaseous disk and eventually trigger star formation every 0.1--1 Gyr, compatible with the estimated age of P3.
This scenario predicted a present-day (i.e., since the starburst of P3) accumulation of $\sim10^4~\rm M_\odot$ of cold (molecular) gas in the M31 nucleus.
However, recent CO observations have placed an upper limit on the molecular gas mass as low as $200~\rm M_\odot$ \citep{2025A&A...693A..24M}, contradicting the prediction of \citet{2007ApJ...668..236C}.
This discrepancy not only recalls questions about the origin of P3 but also highlights the need to understand how stellar winds evolve and dissipate within the M31 NSC, ultimately feeding (or starving) M31*. 

The mass loss of an old stellar population primarily originates from the winds of stars on the red giant branch (RGB) and asymptotic giant branch (AGB).
Particularly, a minor population of thermally pulsing AGB (TP-AGB) stars with short lifetimes ($\sim1~\rm Myr$) and strong winds would dominate the mass-loss \citep{2018A&ARv..26....1H}, except for an extremely old population such as that in globular clusters, i.e., age $>10~\rm Gyr$ \citep{2011MNRAS.416L...6M,2015MNRAS.448..502M,2021MNRAS.503..694T}.
It is conceivable that the dynamical evolution of the AGB star winds, typically being cold and slow, would be drastically different from that of the Wolf-Rayet star winds in the vicinity of Sgr A*. 

In this work, we perform hydrodynamical simulations of M31* accreting stellar winds from the NSC to explore the wind-feeding scenario. The remainder of the paper is organized as follows. In Section \ref{sec:sim}, we describe the setup of hydrodynamical simulations on M31* fed by stellar winds. We present the simulation results and compare the synthetic electromagnetic radiation with observations of M31* in Section \ref{sec:result} and Section \ref{sec:synthesis}, respectively. We discuss the implications for the wind-feeding scenario of dormant SMBHs in general, as well as other factors that would influence the wind-feeding accretion rate in Section \ref{sec:discussion}. We summarize our study in Section \ref{sec:conclusion}.

\section{Simulation Setup}\label{sec:sim}
\subsection{Governing equations}\label{sec:equation}
We perform our grid-based hydrodynamical simulations with the publicly available code \textsc{pluto} \citep[version 4.4patch2;][]{2007ApJS..170..228M}.
The equations we solved are as follows:
\begin{eqnarray*}
\frac{\partial\rho}{\partial t} + \nabla\cdot(\rho \bm{v}) &= \dot{\rho}_w, \\
\frac{\partial\rho \bm{v}}{\partial t} + \nabla\cdot(\rho \bm{vv}+p\mathbb{I}) &= -\rho\nabla\Phi+\dot{\bm{m}}_w, \\
\frac{\partial E_t}{\partial t} + \nabla\cdot[(E_t+p)\bm{v}] &= -\rho\bm{v}\cdot\nabla\Phi + \dot{\rho}_w\Phi +\dot{E}_w+\dot{Q},
\end{eqnarray*}
where $\rho$ is the mass density, $p$ is the thermal pressure, $\bm{v}$ is the velocity, $E_t=p/(\gamma-1)+1/2\rho v^2$ is the total energy density, $\mathbb{I}$ is the identity matrix, $\gamma=5/3$ is the adiabatic index for an ideal equation of state, and $\Phi$ is the gravitational potential.
$\dot{Q}$ is the term combining radiative cooling and photoheating, according to the lookup table presented by \citet{2020MNRAS.497.4857P} based on the photoionization code CLOUDY \citep{2017RMxAA..53..385F}.
Stellar winds are implemented by the source term, $[\dot{\rho}_w, \dot{\bm{m}}_w, \dot{E}_w]$, representing mass, momentum, and energy input, respectively. 
The physical properties of the stellar winds are detailed in Section \ref{sec:wind}.

\begin{figure*}[!hbpt]
	\includegraphics[width=0.45\textwidth, height=0.36246622212717977\textwidth]{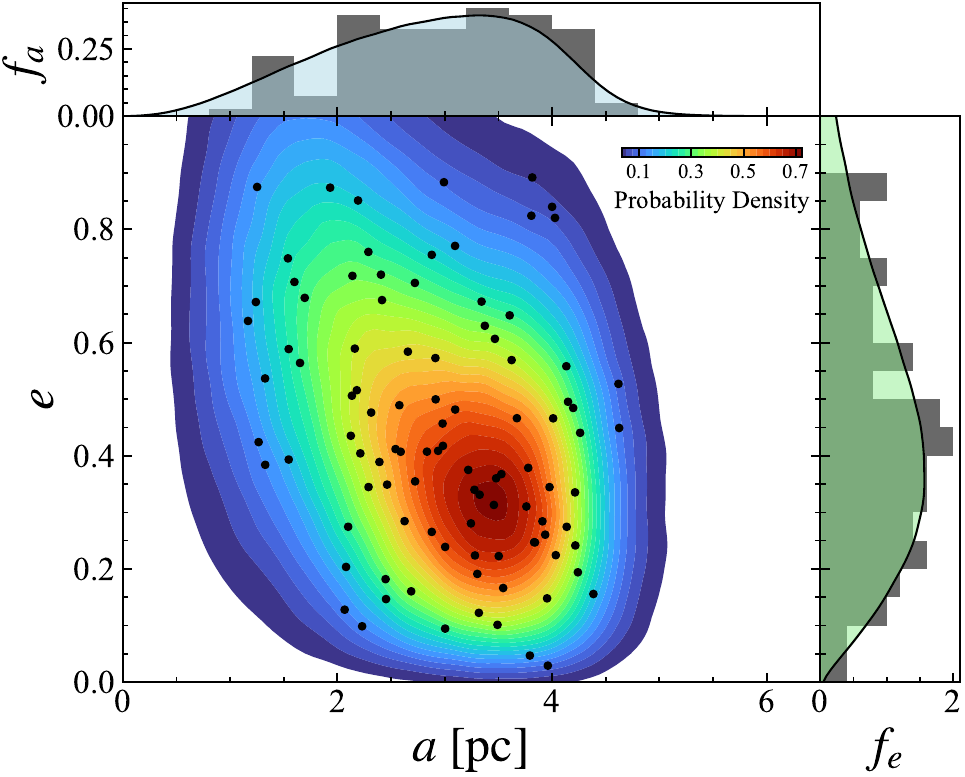}
	\includegraphics[width=0.5\textwidth, height=0.333318\textwidth]{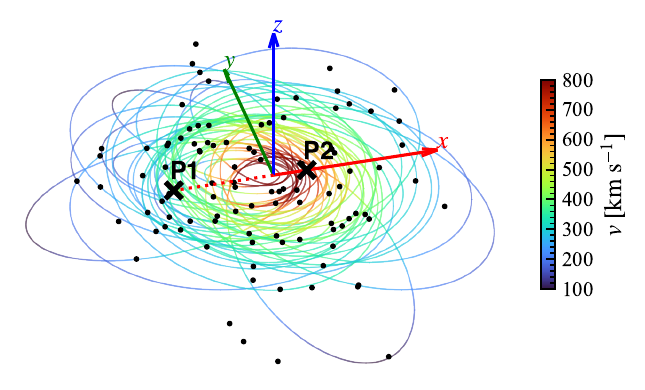}	
	\caption{{\it Left}: Orbital distribution of the \citetalias{2003ApJ...599..237P} NSC model (contours) and the randomly sampled AGB stars (black dots). The side panels display the distributions of the semi-major axis and eccentricity for the NSC model (curve) and randomly sampled AGB stars (histogram). {\it Right}: Initial positions of the AGB stars and trajectories of their eccentric orbits adopted in the simulation. 
    The colors of the trajectories indicate orbital velocities. 
    The $x$-axis is aligned with the P1--P2 direction and the $x-y$ plane corresponds to the mean orbital plane. 
    \label{fig:sample}}
\end{figure*}

We adopt a uniform Cartesian grid covering $16\times16\times8~\rm pc^3$ with $384\times384\times192$ cells, corresponding to a physical resolution of 0.04 pc.
M31* is located at the center of the grid, $(0, 0, 0)$, and the simulation domain encompasses the extent of the NSC. The +$x$-axis is aligned with the P1--P2 direction and the $x$-$y$ plane corresponds to the mean orbital plane, with the +$z$-axis aligned with the mean angular momentum (Figure \ref{fig:sample}).
All three simulations are run for 2.0 Myr (see Section \ref{sec:result} and Section~\ref{subsec:stellarprocess} for justifications), with a timestep of $\sim1$ yr, sufficiently small compared to the orbital period of $10^4-10^5$ yr of the stars.

\subsection{Stellar winds \& orbits}\label{sec:wind}

To estimate the mass-loss from the TP-AGB stars in the NSC, we refer to \textsc{parsec+colibri} stellar isochrones\footnote{generated and interpolated by web interface CMD at \url{https://stev.oapd.inaf.it/cgi-bin/cmd}} \citep{2012MNRAS.427..127B,2013MNRAS.434..488M} to predict the number and average mass-loss rate of TP-AGB stars.
We adopt the Kroupa canonical initial mass function \citep{2001MNRAS.322..231K, 2002Sci...295...82K}, and assume a metallicity of $\rm [M/H]=0.3$, consistent with the metal-rich stellar population in the inner acrsecs of M31, as measured by \citet{2010A&A...509A..61S} using long-slit spectroscopy.
For a total stellar mass of $2\times10^7~\rm M_\odot$ in the NSC and a simple stellar population with an age of $4-12~\rm Gyr$, the isochrones predict the presence of 60--400 TP-AGB and a total mass-loss rate of $(2-14)\times10^{-5}~\rm M_\odot~yr^{-1}$, both decreasing with increasing stellar age.
In our simulation, we adopt a representative case corresponding to an $\sim8~\rm Gyr$ stellar population, with a TP-AGB number of 100, and total mass-loss rate of $4\times10^{-5}~\rm M_\odot~yr^{-1}$, .
The time-averaged mass-loss rate for individual TP-AGB stars is thus $4\times10^{-7}~\rm M_\odot~yr^{-1}$. 
For simplicity, we assume a constant mass-loss rate, although TP-AGB winds can exhibit significant variability during its lifetime.
However, since we are modeling the collective effect of a large number of such stars that would at least partially smooth out the variability of individual sources, we consider this approximation reasonably captures their average behavior.
We neglect the contribution from RGB stars, which are more numerous ($\sim10^5$) but have a much lower value of mass loss rate ($\sim10^{-11}-10^{-10}~\rm M_\odot~yr^{-1}$).
We note that the well studied open cluster NGC\,6791 having an age of $\sim 8~\rm Gyr$ and a metallicity of $\rm [Fe/H]\sim0.3$ \citep{2021A&A...649A.178B} can be regarded as an age-metallicity analogue to the NSC of M31.
Asteroseismology of NGC\,6791 suggested that RGB winds would only contribute to about 15\% of the total mass-loss \citep{2012MNRAS.419.2077M}, which supports the assumed dominance of AGB winds on the stellar mass-loss in the NSC.
Although RGB winds typically have speeds of 20--40 $\rm km~s^{-1}$ \citep{2018ApJ...869....1R, 2024ApJ...967..120W}, several times faster than the massive, slow winds of AGB stars with $v_w\sim10~\rm km~s^{-1}$ \citep{2018A&ARv..26....1H}, the overall velocities of the winds are actually dominated by the orbital motion of the stellar disk at $\sim400~\rm km~s^{-1}$ (see Figure~\ref{fig:sample}).
Therefore, the momentum and kinetic energy contribution from AGB winds remains predominant.
Additionally, we also neglect the stellar winds from the 100--200 Myr population in P3, given that A-type stars are expected to have negligible mass loss and individual B-type stars, if present (later than B5V; $T_{\rm eff}<15000~\rm K$), would have mass-loss rate of $\lesssim6\times10^{-11}~\rm M_\odot~yr^{-1}$ according to the mass-loss recipe for OB stars \citep{2001A&A...369..574V}.
Furthermore, the expected number of TP-AGB stars associated with P3 is only $\approx0.1$, rendering their contribution negligible.

We assume that the motion of each AGB star, like all other old stars in the NSC, follows Keplerian orbits about M31* as modeled by \citet[][hereafter PT03]{2003ApJ...599..237P}, according to which the semi-major axis has a mean value of 3.3 pc, the eccentricity has a mean value of 0.35, and the orbital plane has a mean inclination angle of $54^\circ$.
For each of the 100 AGB stars, we randomly sample the orbital parameters from the \citetalias{2003ApJ...599..237P} model (see illustration in Figure~\ref{fig:sample}).
The sampled AGB stars have a typical pericentric distance of 2 pc, with the most eccentric orbit reaching a minimum distance of 0.16 pc from M31*.
The instantaneous positions and velocities of the AGB stars in the simulation are then determined by solving the Kepler equation.
For each star, we adopt a wind injection radius of 4 cells (0.16 pc) and implement the winds via source term as described in Section \ref{sec:equation}. 
Specifically, the stellar winds are implemented with a temperature of 3000 K and a velocity vector combining the instantaneous orbital velocity (Figure~\ref{fig:sample}) and an isotropic wind velocity of $10~\rm km~s^{-1}$ \citep{2018A&ARv..26....1H}.
In essence, we simulate the wind injection on a star-by-star basis, an approach similar to the simulations for the Wolf-Rayet stars orbiting Sgr A* \citep[e.g.][]{2006MNRAS.366..358C,2018MNRAS.478.3544R}, except that our stellar orbits are determined in a statistical way. 

\subsection{Gravitational potential}\label{gravity}
In our {\tt Fiducial} simulation, the gravitational potential consists of a point mass representing M31* and an extended component representing the eccentric stellar disk, i.e., the NSC. 
As for the former, we adopt a point-mass Keplerian potential with a mass of $M_\bullet=1.0\times10^8~\rm M_\odot$ \citepalias{2003ApJ...599..237P}, which is 
consistent with the dynamical measurement within uncertainties  \citep{2005ApJ...631..280B}. 
As for the latter, the potential is calculated from the density distribution of \citetalias{2003ApJ...599..237P} model by solving the Poisson equation and normalized to a total mass of $2.0\times10^7~\rm M_\odot$.
For simplicity, the eccentric stellar disk is assumed to have no precession, 
which is consistent with the very low precession rate of $\Omega_{\rm P}=0.0\pm3.9~\rm km~s^{-1}~pc^{-1}$ (or $0^\circ\pm0.23^\circ$ kyr$^{-1}$) inferred from near-infrared spectroscopic observations \citep{2018ApJ...854..121L}.
The NSC with its non-axisymmetric potential could enhance the accretion rate onto M31* by two effects, namely, raising the binding energy and driving an inflow \citep{2010MNRAS.405L..41H}. 
Therefore, we also perform a {\tt PointMass} simulation without the NSC potential included (other conditions remain the same as the {\tt Fiducial} simulation) to demonstrate the influence of the NSC potential.

\subsection{External inflow}\label{sec:inflow}
An external inflow, when exists, can also serve as a potential fueling source in addition to the stellar winds from the NSC \citep[e.g.,][]{2023ApJ...953..109A}.  
Gas residing on 10--100 pc scales may be funneled into the central parsecs due to nested bars, magnetic fields, external perturbation, or other mechanisms.
Such an inflow can in fact overwhelm the NSC in terms of fueling the SMBH, a regime beyond the scope of the present study. 
To investigate how a moderate external inflow may affect the stellar wind accretion onto M31*, we perform an additional simulation named {\tt Inflow}, in which an isotropic inflow with a mass rate of $\dot{M}_{\rm inflow}=1.6\times10^{-5}~\rm M_\odot~yr^{-1}$ (i.e. 40\% of the total stellar mass loss in the NSC), a gas temperature of $10^5$ K and an initial speed of 100 km~s$^{-1}$ is implemented, while other conditions remain the same as the {\tt Fiducial} simulation. 
This assumed inflow also effectively absorbs potential stellar winds injected from the large-scale bulge of M31 and/or any gas previously ejected from the NSC but failing to escape to infinity. 
We neglect any net angular momentum that might be carried by the inflow in reality, noting that the assumed spherical symmetry for the inflow should be a good approximation given the small physical scale of the NSC.

\subsection{Initial \& boundary conditions}\label{sec:conditions}
For simplicity, we set the initial condition by assuming a uniformly distributed gas with a number density of $10^{-4}~\rm cm^{-3}$ and a temperature of $10^4~\rm K$.
We find that this low density ambient gas will be rapidly replaced by the stellar winds, which guarantees the dominant role of the stellar winds in the simulation domain.

For the boundary condition, we adopt the {\it outflow} condition that additionally prevents any unphysical inflow for the {\tt Fiducial} and {\tt PointMass} simulations; for the {\tt Inflow} simulation, the boundary condition is set to be the external inflow described in Section \ref{sec:inflow}.
Furthermore, to mimic the accretion onto M31*, gas in the central $2^3$ cells are ``removed" by reset the density and pressure to a substantially small value after every time step in the simulation.
This gives an effective accretion radius of $r_{\rm acc}=8700~r_{\rm g}$ where $r_{\rm g} = GM_{\bullet}/c^2$ is the gravitational radius of M31*.

\begin{figure*}[hbpt]
	\centering
	\includegraphics[width=0.8\linewidth]{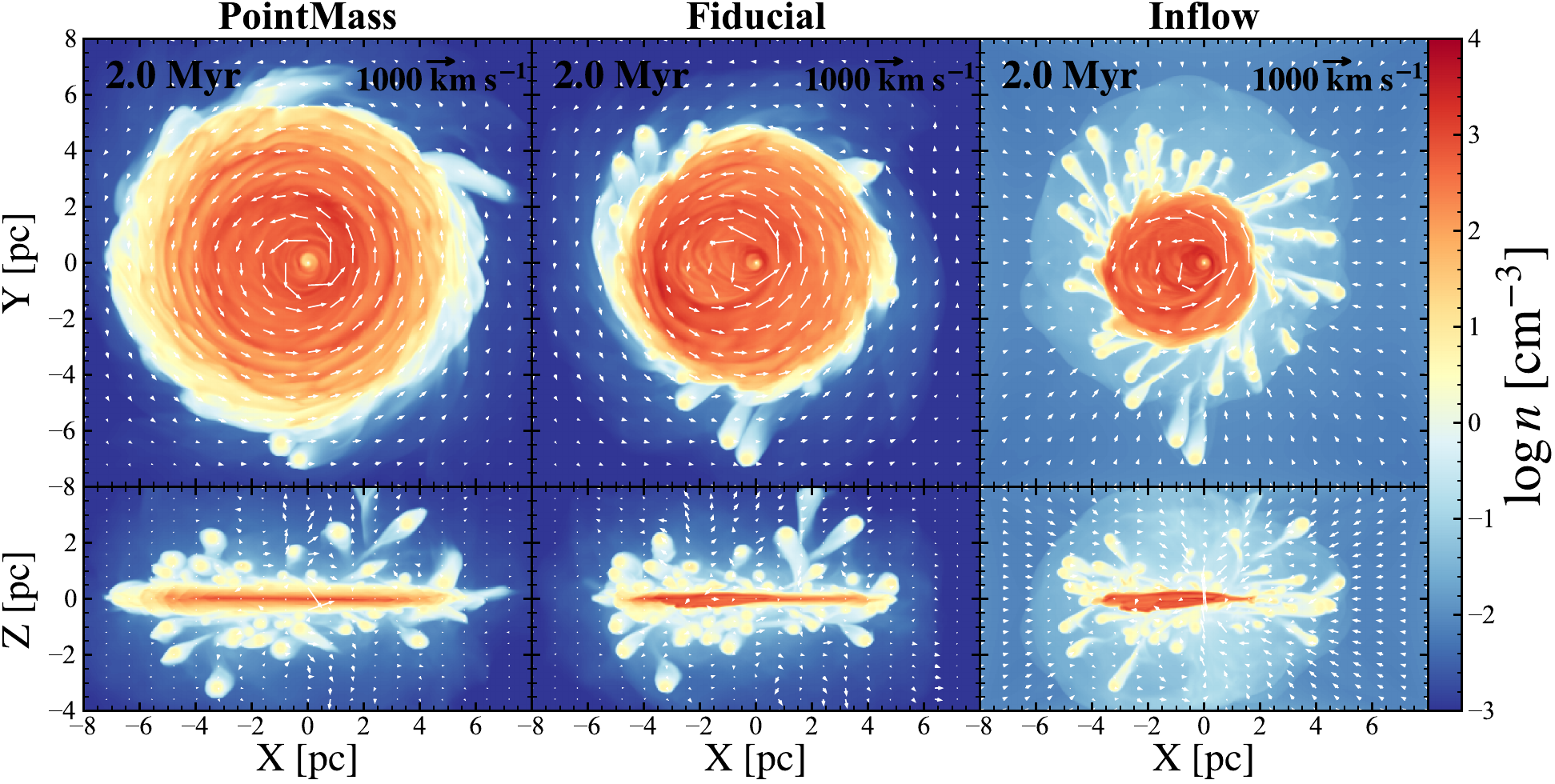}
	\includegraphics[width=0.8\linewidth]{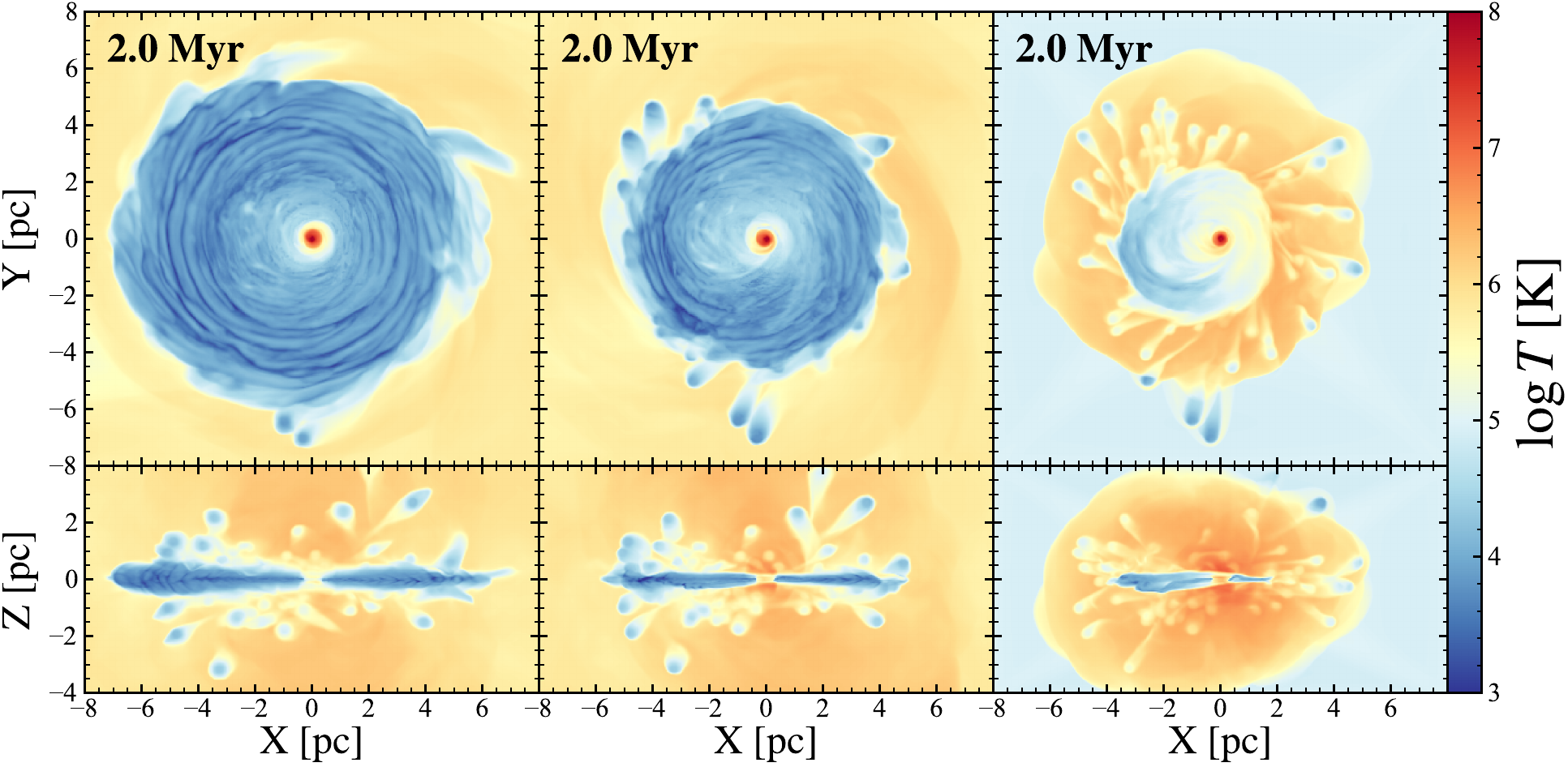}	
	\caption{Mass-weighted density ({\it top}) and temperature ({\it bottom}) distributions for the final snapshot at 2.0 Myr. From left to right, the panels correspond to the {\tt PointMass}, {\tt Fiducial}, and {\tt Inflow} simulations. The distributions are shown as projections onto the mean orbital plane ($x$-$y$ plane) and the plane perpendicular to it ($x$-$z$ plane). Arrows indicate the gas velocity. Wind sources are depicted as blobs in both panels, while the adjacent filament-like structures alongside the blobs correspond to tidally disturbed winds. An animation of the {\tt Fiducial} simulation is available at \url{https://youtu.be/tcKcvro5Fek}. \label{fig:rho_tem}}
\end{figure*}

\section{Simulation Results}\label{sec:result}
In all three simulations, as the stellar winds are continuously injected into the simulation domain, essentially all wind materials are trapped by the gravitational potential, initially retaining the orbital angular momentum of the parent stars and inevitably colliding with each other. 
Eventually, a dense ($n \sim 10^2-10^4\rm~cm^{3}$) and cool ($T \sim 10^{3}-10^5\rm~K$) gas disk builds up around the mean stellar orbital plane, which can be understood as the joint effect of a well-aligned parent angular momentum and strong radiative cooling.
Surrounding the disk is a low-density ($n \sim 10^{-4}-10^{-3}\rm~cm^{3}$), high-temperature ($T \sim 10^{6-7}\rm~K$) halo, which is due to effective thermalization of the kinetic energy of a small fraction of the wind material.
This two-component configuration is markedly different from the case of Sgr A*, where the high-speed ($\sim10^3\rm~km~s^{-1}$) winds of the Wolf-Rayet stars thermalize themselves into a complex network of very hot ($\sim 10^{7-8}$ K) gas via mutual collisions and strong shocks \citep{2006MNRAS.366..358C, 2018MNRAS.478.3544R}.
 Notably, a cold disk of $10^4~\rm K$ has been discovered embedded within the hot plasma in the immediate vicinity of Sgr A* \citep{2019Natur.570...83M}, which to some extent resembles the two-component configuration in our simulation.
Recent simulations suggest that radiative cooling with certain chemical compositions may play an important role in the formation of this disk around Sgr A* \citep{2024ApJ...974...99B,2025A&A...693A.180C}.
Only the innermost region ($\lesssim$ 0.2 pc) is heated to very high temperatures ($T \sim 10^{7-8}\rm~K$) due to the deep gravitational potential of M31*, where a substantial fraction of the gas is accreted into the central sink by design.

The cool disk continues to grow as the stellar winds are continuously injected. By 2.0 Myr, we observe that a quasi-steady state, indicated by a stable disk-halo configuration and a flattened accretion rate, is reached in all three simulations. At this point, a total of $80~\rm M_\odot$ wind material has been injected into the simulation domain.
This time span is consistent with the expected lifetime of the TP-AGB phase with strong mass loss, and in rough agreement with the expected interval for explosive events within NSCs, such as Type Ia supernovae \citep[$\sim1~\rm Myr$;][]{2006ApJ...648..868S} or AGN outbursts \citep[possibly $\sim0.5~\rm Myr$ in M31;][]{2019ApJ...885..157Z}, which could significantly alter the properties of, or even completely disrupt, the wind-fed accretion flow. We note that stellar tidal disruption events (TDEs) could occur on an even shorter timescale ($\sim10^{4-5}$ yr), in view that the mass of M31* ($\sim10^8\rm~M_\odot$) is near the threshold below which a normal star would be tidally disrupted instead of directly swallowed by the SMBH \citep{2021ARA&A..59...21G}.
The effect of such explosive events is reserved for a future study.

In Figure \ref{fig:rho_tem}, we show the face-on and edge-on views of the density and temperature distributions at the final snapshot ($t$ = 2 Myr) of the three simulations.
The cool disk and the hot halo are clearly visible, both constantly disturbed by the windy stars. 
The gas disk has a radius of 3--6 pc, similar to the size of the NSC, and is somewhat lopsided due to the eccentric stellar distribution.
At 2.0 Myr, the accumulated mass of the disk (defined here for cells with a temperature $< 2\times10^4$ K) is about $65~\rm M_\odot$, $40~\rm M_\odot$, and $25~\rm M_\odot$ in the {\tt PointMass}, {\tt Fiducial}, and {\tt Inflow} simulation, respectively. 

\begin{figure*}
    \centering
    \includegraphics[width=0.95\textwidth]{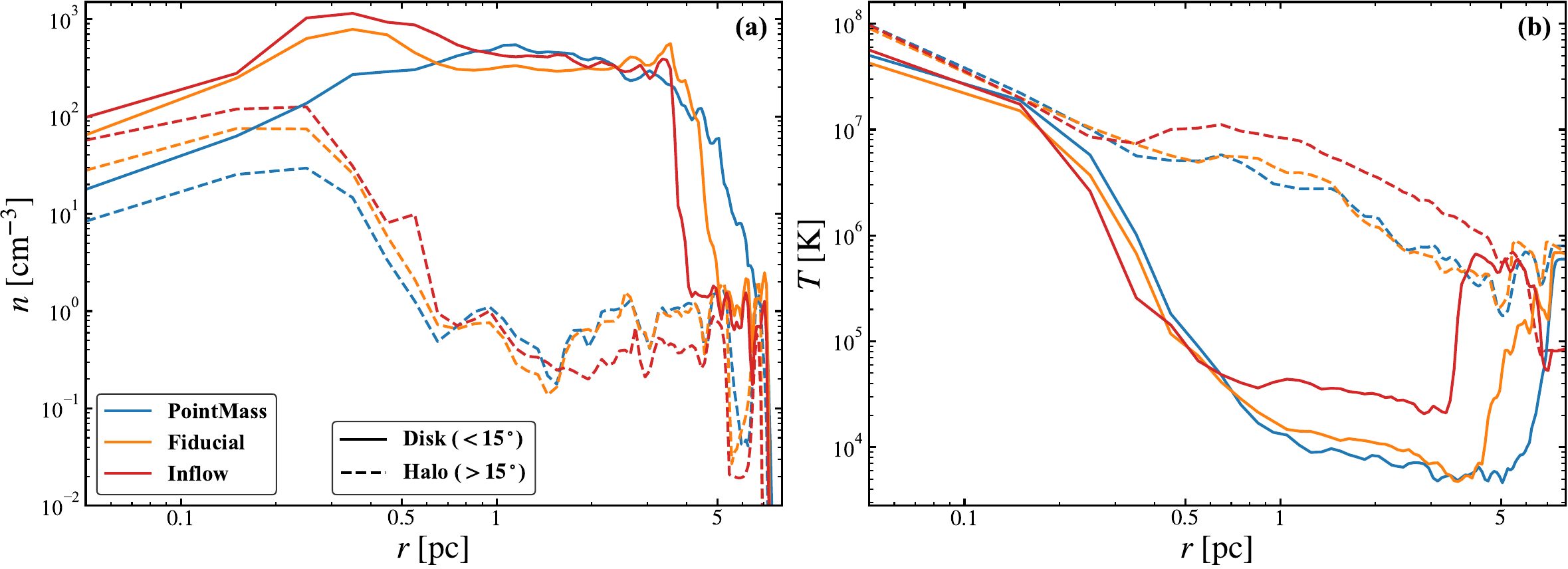}
    \caption{Density ({\it left}) and temperature ({\it right}) radial profiles in the ``disk" (solid curves) and ``halo" (dashed curves) regions. The two regions are defined by the elevation angle relative to the disk plane ($x\text{-}y$ plane). \label{fig:profile}}
\end{figure*}

The wider, less dense but more massive disk in the {\tt PointMass} simulation is due to the shallower gravitational potential compared to the {\tt Fiducial} case.
Uniquely seen in the {\tt Inflow} simulation is a contact discontinuity at a radius of $\sim$4--6 pc, which is the result of a bow shock by the spherical inflow. 
The inflow compresses the disk and the halo, raising the mean density of both components. This compression also causes a higher degree of lopsidedness of the disk.
The inflow is largely held by the pressure of the stellar winds, but a small fraction of the inflow material  can actually reach the vicinity of the SMBH preferentially along the polar regions.
Figure~\ref{fig:profile} shows the azimuthally-averaged gas density and temperature distributions respectively for the disk and halo regions (separated by a polar angle of 15$^\circ$), which further reflects the aforementioned aspects.
Interestingly, the distributions are qualitatively similar to the wind-feeding simulation on Sgr A* by \citet{2025A&A...693A.180C} (see their Figure 4), suggesting a possibly generic outcome of wind-fed accretion. 
We also note that the presence of the gas disk would have negligible impact on the evolution and orbital dynamics of stars crossing it: the ram pressure exerted is $\lesssim10^{-6}$ of the threshold required to strip even the most tenuous stellar envelopes, such as those of AGB stars, and the induced velocity change is $\lesssim10^{-4}$ of typical orbital speeds.

\begin{figure}
	\centering
	\includegraphics[width=0.45\textwidth]{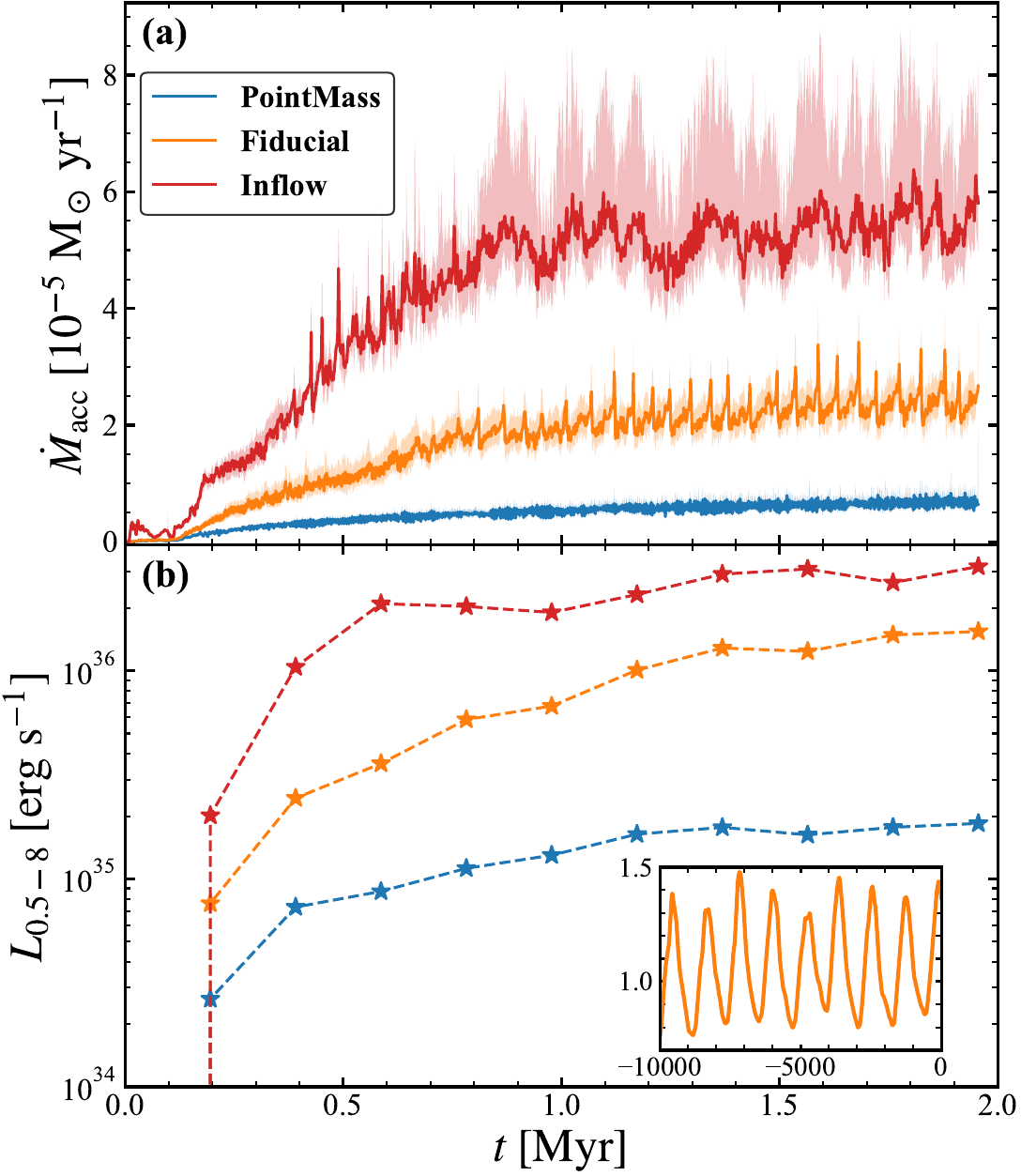}
	\includegraphics[width=0.44\textwidth]{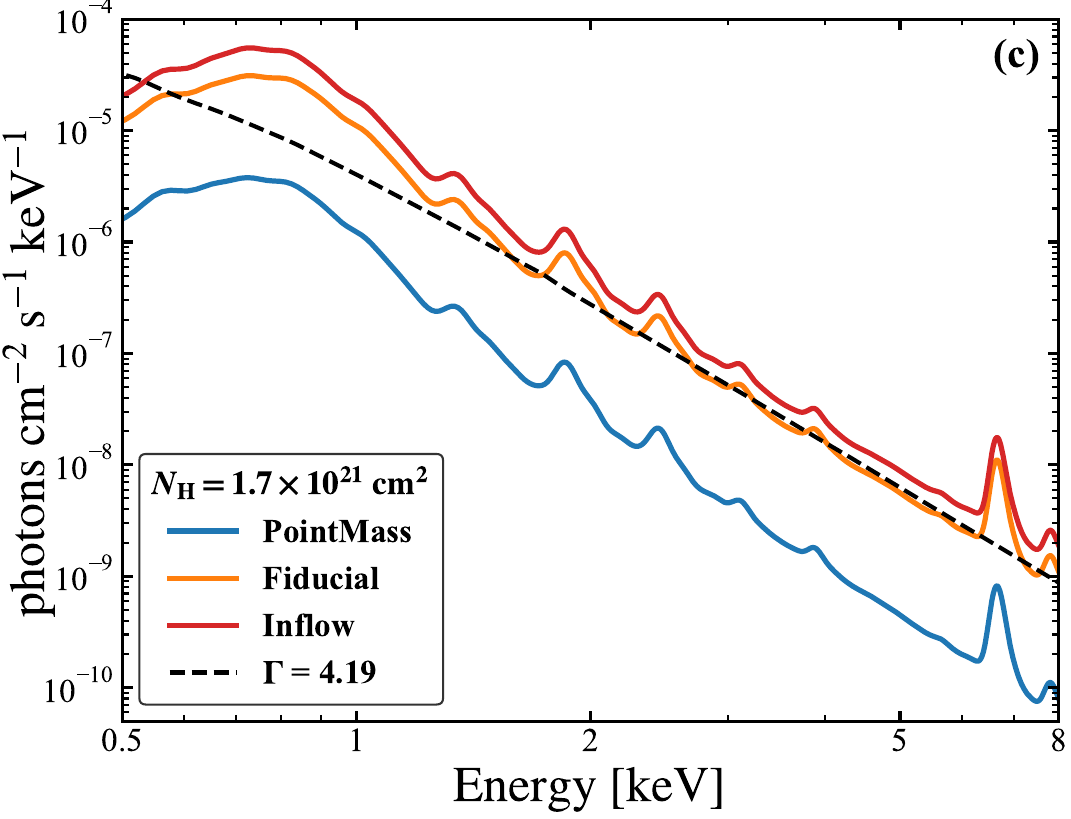}
	\caption{({\it a}): Evolution of the accretion rate. The evolutionary curve is binned in 1000 yr for better trace of the trend. The shaded region represents the minimum and maximum in each bin. ({\it b}): Evolution of the predicted 0.5--8 keV X-ray luminosity, sampled every 0.2 Myr. The inset plots a zoom-in view of the light curve (in units of $10^{36}~\rm erg~s^{-1}$) over the last $10^4~\rm yr$ of the {\tt Fiducial} simulation. ({\it c}): The synthetic X-ray spectra at the snapshot of 2.0 Myr. A foreground absorption of $N_{\rm H}=1.7\times10^{21}~\rm cm^2$ is adopted and the spectra are convolved with the {\it Chandra}/ACIS response matrix with an energy resolution of $\sim100~\rm eV$. The black dashed line represents a power-law spectral analogue, which has a photon index of 4.19. 
    \label{fig:accretion_xray}}	
\end{figure}

 As shown in panel (a) of Figure~\ref{fig:accretion_xray}, the wind-feeding accretion rate at $r_{\rm acc}\sim10^4~r_{\rm g}$ linearly increases with the simulation time in the first Myr and eventually flattens at $6.8\times10^{-6}~\rm M_\odot~yr^{-1}$, $2.4\times10^{-5}~\rm M_\odot~yr^{-1}$, and $5.4\times10^{-5}~\rm M_\odot~yr^{-1}$ for the {\tt PointMass}, {\tt Fiducial}, and {\tt Inflow} simulation, respectively, or 17\%, 60\% and 135\% of the total stellar mass injection rate.  
The non-axisymmetric potential of the NSC induces a 3.5 times higher accretion rate.
The external inflow also significantly increases the accretion rate by both adding direct fuel to the accretion flow and perturbing the gas disk.
That a substantial fraction of the injected wind materials is accreted onto M31* is markedly different from the case of Sgr A*, where the majority of the Wolf-Rayet star winds escape as a bulk outflow after mutual collision and rapid thermalization, and only a minor fraction of the wind material is ultimately accreted by Sgr A* \citep{2018MNRAS.478.3544R}.
Notably, the accretion rate shows frequent fluctuations with a $\sim 50\%$ amplitude, which is mainly driven by the influence of those AGB stars moving near the pericenter, with orbital periods of $\sim10^4~\rm yr$.
Similarly, significant fluctuations in accretion rates have also been found in the wind-fed accretion simulations of Sgr A* \citep[e.g.,][]{2006MNRAS.366..358C, 2018MNRAS.478.3544R}.

\section{Comparison with observations}\label{sec:synthesis}
\subsection{X-ray emission}
\label{subsec:X-ray}
To facilitate a quantitative comparison with the activity of M31* that is directly traced by X-ray emission \citep{2011ApJ...728L..10L, 2025ApJ...981...50D}, we calculate the thermal X-ray emission of the hot plasma in our simulation domain, following the approach in \citet{2022MNRAS.516.1788S}. 
Specifically, the spectrum in the 0.5--8 keV band is extracted from ATOMDB\footnote{\url{http://www.atomdb.org}}, assuming collisional ionization equilibrium and a twice-solar metallicity consistent with the adopted metal-rich stellar population \citep{2010A&A...509A..61S}.
The corresponding X-ray luminosity, integrated over 0.5--8 keV and sampled every 0.2 Myr, is shown in panel (b) of Figure~\ref{fig:accretion_xray} for all three simulations.
We note that the bulk of this luminosity originates from a radius of 0.3 pc, where the accretion flow is the densest and hottest.
It can be seen that the synthetic X-ray luminosities range between $\sim 2\times10^{35}- 3\times10^{36}~\rm erg~s^{-1}$ and are well correlated with the accretion rate (Figure~\ref{fig:accretion_xray}a).
These values, especially that of the {\tt Fiducial} simulation, are in good agreement with the observed mean luminosity of M31*, which is $2\times10^{35}$ and $4\times10^{36}~\rm erg~s^{-1}$ before and after the 2006 flare, respectively \citep{2011ApJ...728L..10L}.
This is an encouraging result, suggesting that our simulations have captured the basic picture of the accretion flow.

We have sampled the synthetic X-ray luminosity at a much finer interval of about 1 yr in the {\tt Fiducial} simulation to trace the short-term variability. 
As shown in the inset of Figure~\ref{fig:accretion_xray}, the synthetic X-ray luminosity varies between $(0.8-1.5)\times10^{36}\rm~erg~s^{-1}$ on a timescale of $\sim1000~\rm yr$ in a quasi-periodic fashion. 
This behavior can again be understood as
modulation by individual AGB stars approaching the pericenter.
We note that the variation amplitude is 
similar to that observed for M31* on a weekly to monthly cadence, except for a couple of large flares which exhibited a much greater flux variation \citep{2011ApJ...728L..10L, 2025ApJ...981...50D}. A more meaningful comparison, however, is currently prevented by the sparse observations and limitation in the temporal/spatial resolution in our simulations. 

We further show in panel (c) of Figure~\ref{fig:accretion_xray} the synthetic X-ray spectra from the hot plasma, which consist of a bremsstrahlung continuum and numerous ionic emission lines.
For comparison, the spectra are in broad agreement with synthetic X-ray spectra of shock-heated and thermalized gas feeding Sgr A*, calculated using a similar approach \citep{2017MNRAS.464.4958R}.
However, at a moderate spectral resolution and low signal-to-noise ratio, typical of the existing {\it Chandra} observations, the thermal characteristics of the spectrum are not apparent. 
\citet{2011ApJ...728L..10L} used a power-law model to characterize the observed spectral hardness ratio and inferred a photon-index of 1.8 (with substantial uncertainty), which is significantly harder than the synthetic X-ray spectrum (close to having a photon-index of 4.2 if mimicked by a power-law; Figure~\ref{fig:accretion_xray}c).
We note that this discrepancy might be alleviated in a higher-resolution simulation zooming into $\sim10-100~r_{\rm g}$, where the temperature of the accretion flow can reach $10^{8-9}~\rm K$ and produce much harder X-ray photons, which are neglected in our current simulations.
Therefore, it is safe to conclude that the wind-feeding scenario can provide a reasonable explanation for the observed X-ray emission from M31*.

\subsection{Cold gas}
\label{subsec:CO}
Regarding the gas reservoir in the NSC, the {\tt Fiducial} simulation predicts the formation of a 40 $\rm M_\odot$ cool gas disk within 2 Myr, which spreads over an extent of a few parsecs (Figure~\ref{fig:rho_tem}).  
In principle, the coldest gas in this disk could be detected via high-resolution observations of CO emission lines.
Based on NOEMA and ALMA CO observations, \citet{2025A&A...693A..24M} reported an upper limit of 195 $\rm M_\odot$ for molecular gas in the central parsec.
While this is consistent with the predicted mass of the cool gas disk, it implies that much more sensitive CO observations are needed to detect the corresponding CO emission.
On the other hand, a substantially more massive disk of approximately $10^4~\rm M_\odot$ is anticipated under the scenario proposed by \citet{2007ApJ...668..236C}, resulting from gravitationally bound stellar mass loss following the formation of P3 (100-200 Myr ago). 
This can be simply scaled to $\sim 10^2~\rm M_\odot$ accumulated on a time interval of 2 Myr, which is compatible with the prediction of the simulations.   

However, as noted in Section~\ref{sec:result}, explosive events including Type Ia supernovae and AGN outburst are expected to occur on a Myr timescale, which could interrupt or halt the growth of the gas disk.
Moreover, in terms of explaining the quiescence of M31* and the observed X-ray emission (Section~\ref{subsec:X-ray}) and achieving consistency with the existing $\rm H\alpha$ measurement (see Section~\ref{subsec:optical} below), there was unlikely a prolonged accumulation of cool gas from the stellar winds since the last star-forming episode 100--200 Myr ago. 
More severely, with the estimated initial stellar mass of a few $10^3\rm~M_\odot$ of the starburst in P3 \citep{2005ApJ...631..280B}, an initial molecular gas of a few $10^4\rm~M_\odot$ is required even for an optimistic star formation efficiency of 10\%.
Therefore, the formation of P3 is rather unlikely to be realized in the scenario proposed by \citet{2007ApJ...668..236C}, and we speculate that a strong inflow was in place to bring about the necessary fueling for the mini-starburst in P3. 

\begin{figure*}
	\centering
	\includegraphics[width=\textwidth]{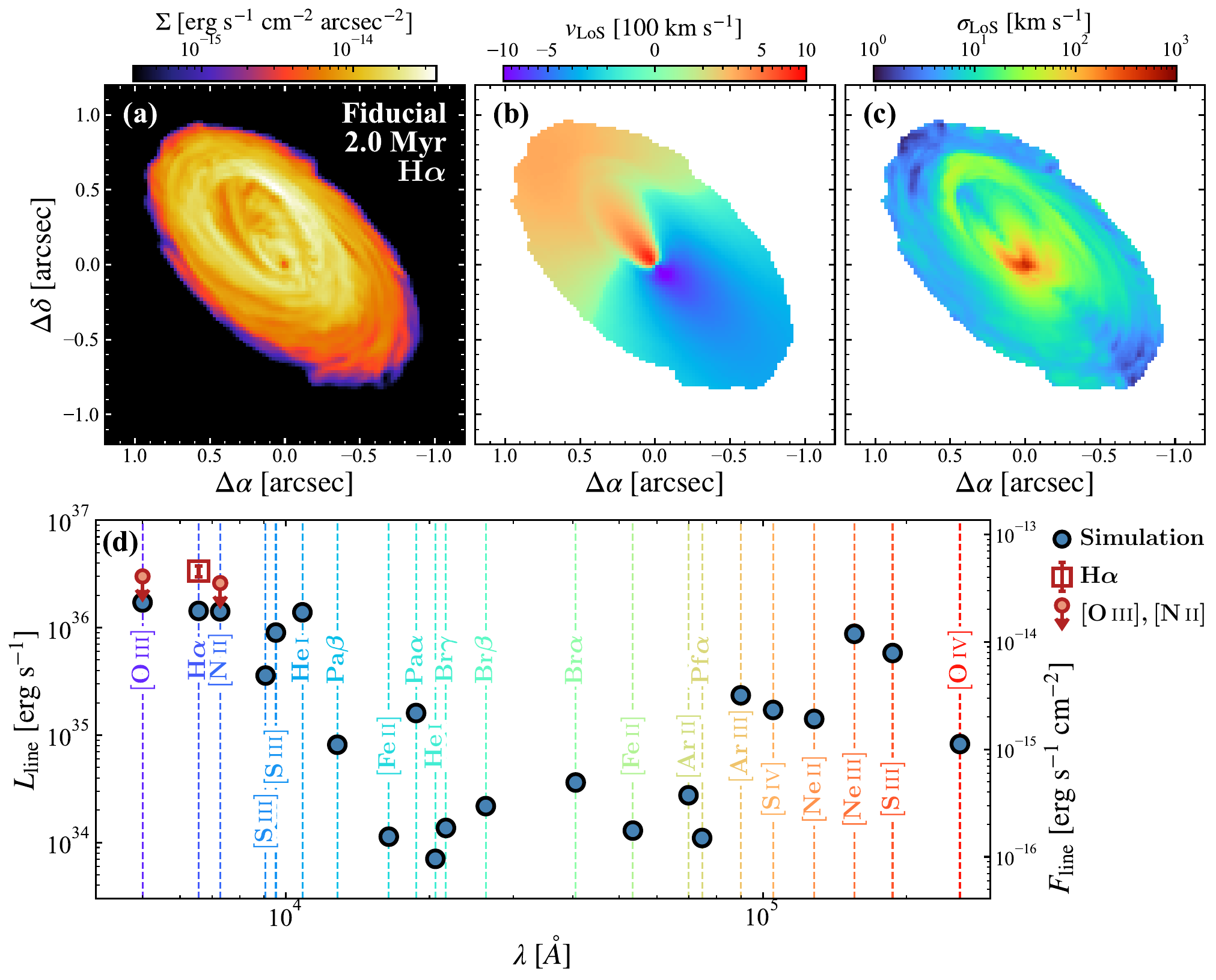}
        \caption{Prediction for optical and infrared emission lines in the {\tt Fiducial} simulation at 2.0 Myr. ({\it a}): synthetic \ha surface brightness map projected on the sky plane. ({\it b}) and ({\it c}): \ha flux-weighted line-of-sight velocity and velocity dispersion maps. In the top three panels, the maps are binned at an angular resolution of 0\farcs022 (equivalent to 2 cells) and the coordinates are relative to M31*. ({\it d}): Total line luminosity ($L_{\rm line}$) of prominent lines in the optical and infrared bands. $F_{\rm line}$ is the line flux for M31's distance of 780 kpc. The \ha measurement (square with error bar; \citealp{2013ApJ...762L..29M}) and  3-$\sigma$ upper limits of [O\,{\sc iii}]$\lambda5007$ and [N\,{\sc ii}]$\lambda6583$ (arrows) of the M31 nucleus are plotted for comparison. The position of [N\,{\sc ii}] is slightly shifted along the horizontal axis for clarity.
        \label{fig:line_emission}}
\end{figure*}

\subsection{Optical \& infrared emission lines}
\label{subsec:optical}
In our simulations, we have neglected potential photoionization, which, in the absence of a luminous AGN, is not expected to significantly affect the gas dynamics. However, photoionization by the strong stellar radiation field in the M31 nucleus might still significantly affect the ionization state of the gas, especially that of the cool disk.
Indeed, despite the lack of young massive stars in the M31 bulge, \citet{2020ApJ...905..138L} found that the far-ultraviolet (UV) intensity in the central hundred parsecs of M31 is $\gtrsim100$ times stronger than the stellar radiation field in the solar neighborhood. 
This high UV flux is originated from hot low-mass evolved stars \citep{2012ApJ...755..131R}, particularly post-AGB stars, which have a surface temperature of $\sim10^5$ K and could serve as the dominant photoionizing source in the central region of M31 (Z.-N. Li et al. in preparation). 

Therefore, we post-process the final snapshot of the {\tt Fiducial} simulation with the photoionization code CLOUDY \citep[version C23;][]{2023RMxAA..59..327C} to model the ionization state and major emission lines from the gas disk.
We first model the broadband spectra and spatial distribution of two stellar components, namely, the NSC of a 8-Gyr-old population, and an ellipsoidal bulge of a 12-Gyr-old population, following the approach in \citet{2023ApJ...958...89L}.
We then derive the position-dependent radiation field and further calculate the line emission of each cell in our simulation domain via CLOUDY.
We note that the A-type stars in P3 are not numerous nor hot enough ($T\sim10^4~\rm K$) to produce a significant number of ionizing photons, which are thus neglected. Also neglected is the weak radiation from M31*.

The {\it top} panels of Figure~\ref{fig:line_emission} displays the two-dimensional view of intensity, line-of-sight velocity and velocity dispersion of the predicted \ha line. 
Clearly, the \ha emission concentrates in the gas disk and preserves a quasi-Keplerian motion with a line-of-sight velocity up to $\sim1000~\rm km~s^{-1}$. 

Due to the intense UV radiation and the relatively small column density ($\sim$$10^{20}\rm~cm^{-2}$) of the disk, CLOUDY predicts that 64\% of the hydrogen atoms in the disk is ionized, which is consistent with the temperature of the disk concentrating at $10^4$ K in the simulation (Figure~\ref{fig:rho_tem}). This is also consistent with the non-detection of CO (Section~\ref{subsec:CO}).
The total synthetic \ha luminosity of the  disk is $1.4\times10^{36}~\rm erg~s^{-1}$ (plotted in Figure \ref{fig:line_emission}d), 
while \citet{2013ApJ...762L..29M} claimed the presence of an $\rm H\alpha$-emitting nuclear disk with an \ha luminosity of $(3.4\pm0.4)\times10^{36}~\rm erg~s^{-1}$ based on integral-field spectroscopic observation of the Gemini North telescope.
We point out that an accurate measurement of the relatively weak \ha emission line relies on a robust modeling of the underlying stellar \ha absorption line from the old stellar population of the NSC. 
Therefore, given our simplified CLOUDY modeling and the potential bias in the stellar continuum modeling, we do not consider the discrepancy between the observed and predicted \ha luminosities to be severe.
On the other hand, the CLOUDY calculation also predicts line luminosities of $1.7\times10^{36}~\rm erg~s^{-1}$ for [O\,{\sc iii}]$\lambda5007$ and $1.4\times10^{36}~\rm erg~s^{-1}$ for [N\,{\sc ii}]$\lambda6583$ from the photoionized gas disk.
These predictions are consistent with the non-detections by the  Gemini North integral-field spectroscopic observation, which places 3-$\sigma$ upper limits of $3.0\times10^{36}~\rm erg~s^{-1}$ and $2.6\times10^{36}~\rm erg~s^{-1}$, respectively. 
In Figure \ref{fig:line_emission}d, we plot the predicted luminosities of $\rm H\alpha$, [O\,{\sc iii}], [N\,{\sc ii}], and a number of other optical and infrared emission lines from the gas disk, bearing in mind the aforementioned systematics in the CLOUDY calculation. 
Interestingly, many of these emission lines, especially those in the infrared band, are within reach by current facilities such as JWST, which call for future observations that can help validate our simulations and provide useful diagnostics of the accretion process.

\section{Discussion}\label{sec:discussion}
\subsection{Stellar processes in the NSC}
\label{subsec:stellarprocess}
In our simulations, we determine the mass-loss from the old stellar population based on stellar evolution modeling.
However, in the extreme environment of the NSC, characterized by a high stellar density and proximity to the central SMBH (here M31*), stellar dynamical processes and interactions with the SMBH could, in principle, influence the evolution and the mass-loss behavior of the stars \citep[see review by][]{2017ARA&A..55...17A}. 
Here, we discuss the potentially relevant effects.

{\it Stellar collisions.}---Considering the high stellar density and velocity dispersion in the NSC, stellar collisions might destroy the envelope of giant stars before they evolve into the AGB phase. 
This is reminiscent of the observed absence of bright giant stars in the Galactic center, which is suggested to originate from stellar collisions \citep{1999MNRAS.308..257B,2009MNRAS.393.1016D,2023ApJ...955...30R}.
According to \citet{2008gady.book.....B}, the collision time scale $t_{\rm coll}$ for star-star collision can be expressed as 
\begin{eqnarray}
t_{\rm coll}^{-1}&=&4\sqrt{\pi}n\sigma(R_{*}+R_{\rm imp})^2[1+\frac{G(M_*+M_{\rm imp})}{\sigma^2(R_{*}+R_{\rm imp})}],
\end{eqnarray}
where $n\sim10^6~pc^{-3}$ is the mean number density of NSC stars, $\sigma\sim150~\rm km~s^{-1}$ is the velocity dispersion, and $R_*$ ($R_{\rm imp}$) and $M_*$ ($M_{\rm imp}$) are radius and mass of the target (impactor) star.
Since the radius of the star of interest here, i.e., RGB star, would vary significantly throughout its life span, we estimate the time-integrated probability of collision as $p_{\rm coll}=\int t_{\rm coll}^{-1}(t)dt$ following \citet{2009MNRAS.393.1016D}.
Specifically, we assume the target star to be an RGB star with an initial mass of $1.15~\rm M_\odot$, whose evolutionary track is taken from the PARSEC stellar tracks \citep{2015MNRAS.452.1068C}, and the impactor stars to be solar-type stars.
The radius of the RGB star would increase from 2 to 200 $R_\odot$ over its 0.5 Gyr lifetime.
Under these assumptions, the collision probability for a star during the RGB phase is estimated to be 0.09, suggesting that most RGB stars in the stellar disk remain unaffected.
We note that the collision probability could reach as high as 0.5 at the P2 position due to the higher density and velocity dispersion.
However, stars spend a very short time at P2 (the pericenter), limiting the overall impact of collisions.

{\it Tidal spin-up.}---Compared with the destructive stellar collisions, close encounters having a higher occurring frequency would transfer orbital energy and angular momentum to the involved stars.
In the context of the Galactic center, \citet{2001ApJ...549..948A} found that the stochastic tidal spin-up from hyperbolic encounters would cumulatively result in an unusually high rotation rate ($\gtrsim10\%$ of the centrifugal breakup velocity) for the main-sequence stars within the central 0.3 pc.
An atypically high spin may be a distinguished feature for stars secularly evolving in the dense stellar cluster near an SMBH \citep{2005PhR...419...65A}, which is relevant to the NSC of M31. 
However, the evolution of low-mass giant stars is largely insensitive to the initial rotation rate of the main-sequence star, even when the rotation rate reaches 90\% of the breakup velocity \citep{2022A&A...665A.126N}. This insensitivity arises because the expansion of the stellar envelope significantly reduces the angular velocity. Furthermore, the direct tidal spin-up effect on giant stars is weak due to their relatively short lifetimes.
Therefore, the tidal spin-up would have a negligible influence on the mass-loss from AGB stars.

{\it Tidal disruption events.}---TDEs are inherent consequences for the NSC within the sphere of influence of the central SMBH. 
TDEs could strip the puffed, gaseous envelope of the giant stars and the disruption rate for these stars could be higher than that for main-sequence stars due to their larger tidal radii.
\citet{2012ApJ...757..134M} predicted that the relation between the disruption rate $\dot{n}$ scales with tidal radius $r_t$ as $\dot{n}\propto r_t^{1/4}$, and the disruption of giant stars, primarily RGB stars in the case of low-mass stars, would contribute about 10\% of total TDEs.
Given a typical TDE rate on the order of $10^{-4}~{\rm yr^{-1}~gal^{-1}}$, the time scale for an RGB star experiencing a TDE, $\tau=N_{\rm RGB}/(0.1\dot{N}_{\rm TDE})$, is estimated to be $\sim10^{10}~\rm yr$ which is much larger than the lifetime of the RGB phase.
An eccentric stellar disk is possibly subject to an enhanced TDE rate due to secular torques but the TDE rate is suggested to be extremely high only during the very early evolutionary stages and would drastically reduce when the disk stabilizes \citep{2018ApJ...853..141M, 2019ApJ...880...42W}, as for the case of M31.

In summary, although there are various dynamical processes that could influence the stellar evolution of the NSC, we find that the majority of the AGB stars would not be significantly affected by these processes and safely follow the general evolutionary track. It is further noteworthy that the stellar envelope loss due to stellar collisions would likely be retained in the NSC, and ultimately join the gas disk and become fuel of the SMBH. Therefore, we conclude that the assumed uninterrupted stellar wind injection in our simulations is reasonable and the main aspects of the resultant accretion flow should be robust.

\subsection{Numerical resolution}\label{subsec:resolution}
Due to the limited resolution mainly constrained by computational resource, our simulations adopt an effective accretion radius of $r_{\rm acc}\approx8700~r_{\rm g}$ for M31* and neglect the behavior of the accretion flow further inside. 
The choice of $r_{\rm acc}$ certainly affects to some extent the overall dynamics and predicted radiation of the hot accretion flow, as demonstrated in the series simulations by \citet{2018MNRAS.478.3544R} for the case of Sgr A*, but the net effect can only be quantified with higher-resolution simulations, which we defer for future work.  
Qualitatively, the exclusion of hot gas within $r_{\rm acc}$ would result in an underestimate of the synthetic X-ray luminosity and its hardness.
Moreover, the dynamical timescale at the current $r_{\rm acc}$ is 75 yr, which is incompatible to the cadence of X-ray observations of M31*. 
To capture a more realistic variability would require a higher resolution in the vicinity of the SMBH.
To gain insight into the impact of resolution, we have performed a zoom-in simulation starting from the final snapshot of the {\tt Fiducial} simulation with static mesh refinement, effectively reducing the accretion radius to $r_{\rm acc}\approx1000~r_{\rm g}$.
We run the simulation for 0.1 Myr at a substantially reduced time step, which is sufficiently long for the accretion rate and X-ray luminosity to reach a new quasi-steady state.
This refined simulation yields an on-average 4 times higher X-ray luminosity, a harder X-ray spectrum with a pseudo-photon-index of 3.2, 
and a shorter variation timescale of 25 yr.
These results align with expectations, and the increased X-ray luminosity is still compatible with the observed range \citep{2011ApJ...728L..10L}.

\subsection{Implications for dormant SMBHs}
Through dedicated hydrodynamical simulations, we demonstrate that M31* can be efficiently fueled by stellar winds from its surrounding NSC and manifested in multi-wavelength emission in reasonable agreement with existing observations.
Therefore, M31* joins Sgr A* as a firm case in which the exceptionally low accretion state ($L_{\rm bol}/L_{\rm Edd}\sim10^{-9}$) of an SMBH is dictated by stellar wind-feeding, despite the stark difference in the physical properties of the winds and the parent stars. 
It is natural to consider whether the wind-feeding scenario works for
other nearby dormant SMBHs, which typically have a higher Eddington ratio ($L_{\rm bol}/L_{\rm Edd}\lesssim 10^{-3}$) than M31*.
For the more general case where the NSC is nearly spherical and axisymmetric, the behavior of the wind-fed accretion flow can be qualitatively different. 
An axisymmetric potential would be less capable of driving an inflow than a non-axisymmetric potential serving as a torque, as is the case of the M31 NSC.
On the other hand, the net angular momentum of the stellar winds from a spherical, dynamically hot NSC can be lower than that of a disk-shaped NSC.
This may compensate for the lack of the torque and result in a smaller, denser gas disk and thus a higher accretion rate.
Additionally, an NSC could be relatively more massive than the case of M31* (with $M_{\rm NSC}/M_\bullet\approx 0.2$). 
Also the stellar density in the immediate vicinity of the central SMBH can be significantly higher than that of M31*, because a stellar cusp is predicted to form around the SMBH residing in a spherical NSC \citep{1976ApJ...209..214B}.
All the above factors can alter the wind-fed accretion flow and lead to a substantially higher accretion rate, which is expected for many known local SMBHs with a higher Eddington ratio.
Inferring the wind-fed accretion rate in a general context thus calls for systematic simulations of wind-fed accretion.

The existence of a symbiotic hot wind is theoretically and numerically predicted for hot accretion flows onto sub-Eddington SMBHs \citep{2014ARA&A..52..529Y}, which has the potential of not only modifying the accretion process itself but also influencing the host galaxy on larger scales.
Recently, the existence of hot winds has been observationally confirmed by X-ray spectroscopy of nearby LLAGNs (M81*, \citealp{2021NatAs...5..928S}; NGC\,7213, \citealp{2022ApJ...926..209S}).
However, no clear sign of a persistent outflow is seen in our standard simulations. This is likely due to the limited numerical resolution that cannot self-consistently capture the wind launching zone (expected to be $r\lesssim 1000~r_{\rm g}$) and also to the absence of magnetic fields which could help to collimate and launch winds \citep{2015ApJ...804..101Y,2020ApJ...896L...6R}.
Interestingly, in the refined simulation discussed in Section~\ref{subsec:resolution}, we observe a persistent polar outflow with an opening angle of $\approx 30^\circ$ and a radial velocity of $\sim3000~\rm km~s^{-1}$, which could be trace of the putative hot wind. This outflow is most likely driven by the strong gas pressure gradient at small radii.
The mechanical output from the hot winds would resist the accretion flow, especially in the direction far from the disk plane, thus effectively suppressing the accretion rate. A more detailed investigation of this effect in the context of wind-fed accretion is reserved for future work.

Studying the role of NSCs in the accretion onto SMBHs can also complement cosmological simulations.
Due to the limitation of resolution, state-of-the-art cosmological simulations universally adopt a sub-grid model for BH feeding, typically based on {\it Bondi accretion}.
This approach makes SMBH accretion primarily determined by the gas properties within hundreds of parsecs of the SMBH, while neglecting the mass and energy supply from the NSC, which, as demonstrated in the present work, might be significant for regulating accretion onto highly sub-Eddington SMBHs.
Therefore, a systematic study on wind-fed accretion by the NSCs would help improve the sub-grid model of SMBH accretion and the understanding of the co-evolution of SMBHs and their host galaxies over cosmic time.

\section{Conclusions}\label{sec:conclusion}
In this paper, we perform hydrodynamical simulations on M31* fed by stellar mass-loss from the nuclear star cluster, which takes a rather unusual configuration of an eccentric, {apsidally} aligned stellar disk.
The stellar mass-loss are dominated by slow stellar winds from 100 TP-AGB stars on Keplerian orbits around M31*.
We find that M31* can have an accretion rate of $\sim10^{-5}~\rm M_\odot~yr^{-1}$ and an X-ray luminosity of $\sim10^{36}~\rm erg~s^{-1}$ that is well consistent with {\it Chandra} observations.
Both the gravitational potential of the NSC and an assumed external inflow can significantly enhance the accretion rate and hence the X-ray luminosity.
Besides the hot plasma in the vicinity of M31*, the majority of the stellar winds would settle as a gas disk with a mass of $\sim 40~\rm M_\odot$ within $\sim$2 Myr.
The gas disk can be photoionized by the old stellar populations in the NSC and the bulge.
The predicted \ha line emission from the gas disk also aligns with optical spectroscopic observations.
Other recombination and forbidden lines in the optical and infrared bands (Figure \ref{fig:line_emission}) remain to be uncovered to test the wind-feeding scenario and to provide useful diagnostics to the accretion flow and the SMBH itself.
Our work demonstrates that M31* can be fed by stellar winds from the NSC and also highlights the plausibility of the wind-feeding scenario for dormant SMBHs in general.

Inferring the properties of wind-fed accretion onto other quiescent SMBHs is still challenging, as both M31* and Sgr A* exhibit unique characteristics in the dynamics or stellar populations of their NSCs.
To better understand the role of NSCs in feeding SMBHs, further simulations of generic NSC-SMBH systems are required and will be conducted in our future work.

\begin{acknowledgements}
The authors would like to thank Luis C. Ho, Q. Daniel Wang and Feng Yuan for helpful discussions, and Jiachang Zhang and Sumin Wang for assistance with the observational data. This work is supported by the National Natural Science Foundation of China (grant 12225302), the National Key Research and Development Program of China (No. 2022YFF0503402) and CNSA program D050102. Z.N.L. acknowledges support from the East Asian Core Observatories Association Fellowship.
\end{acknowledgements}

\software{PLUTO \citep{2007ApJS..170..228M},
          AtomDB \citep{2012ApJ...756..128F},
          CLOUDY \citep{2023RMxAA..59..327C}
          }

\bibliography{main}{}

\begin{thebibliography}{}
\expandafter\ifx\csname natexlab\endcsname\relax\def\natexlab#1{#1}\fi
\providecommand{\url}[1]{\href{#1}{#1}}
\providecommand{\dodoi}[1]{doi:~\href{http://doi.org/#1}{\nolinkurl{#1}}}
\providecommand{\doeprint}[1]{\href{http://ascl.net/#1}{\nolinkurl{http://ascl.net/#1}}}
\providecommand{\doarXiv}[1]{\href{https://arxiv.org/abs/#1}{\nolinkurl{https://arxiv.org/abs/#1}}}

\bibitem[{{Alexander}(2005)}]{2005PhR...419...65A}
{Alexander}, T. 2005, \physrep, 419, 65, \dodoi{10.1016/j.physrep.2005.08.002}

\bibitem[{{Alexander}(2017)}]{2017ARA&A..55...17A}
---. 2017, \araa, 55, 17, \dodoi{10.1146/annurev-astro-091916-055306}

\bibitem[{{Alexander} \& {Kumar}(2001)}]{2001ApJ...549..948A}
{Alexander}, T., \& {Kumar}, P. 2001, \apj, 549, 948, \dodoi{10.1086/319436}

\bibitem[{{Alig} {et~al.}(2023){Alig}, {Prieto}, {Bla{\~n}a}, {Frischman}, {Metzl}, {Burkert}, {Zier}, \& {Streblyanska}}]{2023ApJ...953..109A}
{Alig}, C., {Prieto}, A., {Bla{\~n}a}, M., {et~al.} 2023, \apj, 953, 109, \dodoi{10.3847/1538-4357/ace2c3}

\bibitem[{{Bahcall} \& {Wolf}(1976)}]{1976ApJ...209..214B}
{Bahcall}, J.~N., \& {Wolf}, R.~A. 1976, \apj, 209, 214, \dodoi{10.1086/154711}

\bibitem[{{Bailey} \& {Davies}(1999)}]{1999MNRAS.308..257B}
{Bailey}, V.~C., \& {Davies}, M.~B. 1999, \mnras, 308, 257, \dodoi{10.1046/j.1365-8711.1999.02740.x}

\bibitem[{{Balakrishnan} {et~al.}(2024){Balakrishnan}, {Russell}, {Corrales}, {Calder{\'o}n}, {Cuadra}, {Haggard}, {Markoff}, {Neilsen}, {Nowak}, {Wang}, \& {Baganoff}}]{2024ApJ...974...99B}
{Balakrishnan}, M., {Russell}, C. M.~P., {Corrales}, L., {et~al.} 2024, \apj, 974, 99, \dodoi{10.3847/1538-4357/ad6866}

\bibitem[{{Bender} {et~al.}(2005){Bender}, {Kormendy}, {Bower}, {Green}, {Thomas}, {Danks}, {Gull}, {Hutchings}, {Joseph}, {Kaiser}, {Lauer}, {Nelson}, {Richstone}, {Weistrop}, \& {Woodgate}}]{2005ApJ...631..280B}
{Bender}, R., {Kormendy}, J., {Bower}, G., {et~al.} 2005, \apj, 631, 280, \dodoi{10.1086/432434}

\bibitem[{{Binney} \& {Tremaine}(2008)}]{2008gady.book.....B}
{Binney}, J., \& {Tremaine}, S. 2008, {Galactic Dynamics: Second Edition}

\bibitem[{{Bressan} {et~al.}(2012){Bressan}, {Marigo}, {Girardi}, {Salasnich}, {Dal Cero}, {Rubele}, \& {Nanni}}]{2012MNRAS.427..127B}
{Bressan}, A., {Marigo}, P., {Girardi}, L., {et~al.} 2012, \mnras, 427, 127, \dodoi{10.1111/j.1365-2966.2012.21948.x}

\bibitem[{{Brogaard} {et~al.}(2021){Brogaard}, {Grundahl}, {Sandquist}, {Slumstrup}, {Jensen}, {Thomsen}, {J{\o}rgensen}, {Larsen}, {Bj{\o}rn}, {S{\o}rensen}, {Bruntt}, {Arentoft}, {Frandsen}, {Jessen-Hansen}, {Orosz}, {Mathieu}, {Geller}, {Ryde}, {Stello}, {Meibom}, \& {Platais}}]{2021A&A...649A.178B}
{Brogaard}, K., {Grundahl}, F., {Sandquist}, E.~L., {et~al.} 2021, \aap, 649, A178, \dodoi{10.1051/0004-6361/202140911}

\bibitem[{{Calder{\'o}n} {et~al.}(2025){Calder{\'o}n}, {Cuadra}, {Russell}, {Burkert}, {Rosswog}, \& {Balakrishnan}}]{2025A&A...693A.180C}
{Calder{\'o}n}, D., {Cuadra}, J., {Russell}, C. M.~P., {et~al.} 2025, \aap, 693, A180, \dodoi{10.1051/0004-6361/202452800}

\bibitem[{{Calder{\'o}n} {et~al.}(2020){Calder{\'o}n}, {Cuadra}, {Schartmann}, {Burkert}, \& {Russell}}]{2020ApJ...888L...2C}
{Calder{\'o}n}, D., {Cuadra}, J., {Schartmann}, M., {Burkert}, A., \& {Russell}, C. M.~P. 2020, \apjl, 888, L2, \dodoi{10.3847/2041-8213/ab5e81}

\bibitem[{{Carson} {et~al.}(2015){Carson}, {Barth}, {Seth}, {den Brok}, {Cappellari}, {Greene}, {Ho}, \& {Neumayer}}]{2015AJ....149..170C}
{Carson}, D.~J., {Barth}, A.~J., {Seth}, A.~C., {et~al.} 2015, \aj, 149, 170, \dodoi{10.1088/0004-6256/149/5/170}

\bibitem[{{Chang} {et~al.}(2007){Chang}, {Murray-Clay}, {Chiang}, \& {Quataert}}]{2007ApJ...668..236C}
{Chang}, P., {Murray-Clay}, R., {Chiang}, E., \& {Quataert}, E. 2007, \apj, 668, 236, \dodoi{10.1086/521018}

\bibitem[{{Chatzikos} {et~al.}(2023){Chatzikos}, {Bianchi}, {Camilloni}, {Chakraborty}, {Gunasekera}, {Guzm{\'a}n}, {Milby}, {Sarkar}, {Shaw}, {van Hoof}, \& {Ferland}}]{2023RMxAA..59..327C}
{Chatzikos}, M., {Bianchi}, S., {Camilloni}, F., {et~al.} 2023, \rmxaa, 59, 327, \dodoi{10.22201/ia.01851101p.2023.59.02.12}

\bibitem[{{Chen} {et~al.}(2015){Chen}, {Bressan}, {Girardi}, {Marigo}, {Kong}, \& {Lanza}}]{2015MNRAS.452.1068C}
{Chen}, Y., {Bressan}, A., {Girardi}, L., {et~al.} 2015, \mnras, 452, 1068, \dodoi{10.1093/mnras/stv1281}

\bibitem[{{Crane} {et~al.}(1993){Crane}, {Cowan}, {Dickel}, \& {Roberts}}]{1993ApJ...417L..61C}
{Crane}, P.~C., {Cowan}, J.~J., {Dickel}, J.~R., \& {Roberts}, D.~A. 1993, \apjl, 417, L61, \dodoi{10.1086/187094}

\bibitem[{{Crane} {et~al.}(1992){Crane}, {Dickel}, \& {Cowan}}]{1992ApJ...390L...9C}
{Crane}, P.~C., {Dickel}, J.~R., \& {Cowan}, J.~J. 1992, \apjl, 390, L9, \dodoi{10.1086/186359}

\bibitem[{{Cuadra} {et~al.}(2006){Cuadra}, {Nayakshin}, {Springel}, \& {Di Matteo}}]{2006MNRAS.366..358C}
{Cuadra}, J., {Nayakshin}, S., {Springel}, V., \& {Di Matteo}, T. 2006, \mnras, 366, 358, \dodoi{10.1111/j.1365-2966.2005.09837.x}

\bibitem[{{Dale} {et~al.}(2009){Dale}, {Davies}, {Church}, \& {Freitag}}]{2009MNRAS.393.1016D}
{Dale}, J.~E., {Davies}, M.~B., {Church}, R.~P., \& {Freitag}, M. 2009, \mnras, 393, 1016, \dodoi{10.1111/j.1365-2966.2008.14254.x}

\bibitem[{{DiKerby} {et~al.}(2025){DiKerby}, {Zhang}, \& {Irwin}}]{2025ApJ...981...50D}
{DiKerby}, S., {Zhang}, S., \& {Irwin}, J. 2025, \apj, 981, 50, \dodoi{10.3847/1538-4357/adb1d5}

\bibitem[{{Feldmeier-Krause} {et~al.}(2015){Feldmeier-Krause}, {Neumayer}, {Sch{\"o}del}, {Seth}, {Hilker}, {de Zeeuw}, {Kuntschner}, {Walcher}, {L{\"u}tzgendorf}, \& {Kissler-Patig}}]{2015A&A...584A...2F}
{Feldmeier-Krause}, A., {Neumayer}, N., {Sch{\"o}del}, R., {et~al.} 2015, \aap, 584, A2, \dodoi{10.1051/0004-6361/201526336}

\bibitem[{{Ferland} {et~al.}(2017){Ferland}, {Chatzikos}, {Guzm{\'a}n}, {Lykins}, {van Hoof}, {Williams}, {Abel}, {Badnell}, {Keenan}, {Porter}, \& {Stancil}}]{2017RMxAA..53..385F}
{Ferland}, G.~J., {Chatzikos}, M., {Guzm{\'a}n}, F., {et~al.} 2017, \rmxaa, 53, 385, \dodoi{10.48550/arXiv.1705.10877}

\bibitem[{{Foster} {et~al.}(2012){Foster}, {Ji}, {Smith}, \& {Brickhouse}}]{2012ApJ...756..128F}
{Foster}, A.~R., {Ji}, L., {Smith}, R.~K., \& {Brickhouse}, N.~S. 2012, \apj, 756, 128, \dodoi{10.1088/0004-637X/756/2/128}

\bibitem[{{Garcia} {et~al.}(2005){Garcia}, {Williams}, {Yuan}, {Kong}, {Primini}, {Barmby}, {Kaaret}, \& {Murray}}]{2005ApJ...632.1042G}
{Garcia}, M.~R., {Williams}, B.~F., {Yuan}, F., {et~al.} 2005, \apj, 632, 1042, \dodoi{10.1086/432967}

\bibitem[{{Gezari}(2021)}]{2021ARA&A..59...21G}
{Gezari}, S. 2021, \araa, 59, 21, \dodoi{10.1146/annurev-astro-111720-030029}

\bibitem[{{Ho}(2008)}]{2008ARA&A..46..475H}
{Ho}, L.~C. 2008, \araa, 46, 475, \dodoi{10.1146/annurev.astro.45.051806.110546}

\bibitem[{{Ho}(2009)}]{2009ApJ...699..626H}
---. 2009, \apj, 699, 626, \dodoi{10.1088/0004-637X/699/1/626}

\bibitem[{{H{\"o}fner} \& {Olofsson}(2018)}]{2018A&ARv..26....1H}
{H{\"o}fner}, S., \& {Olofsson}, H. 2018, \aapr, 26, 1, \dodoi{10.1007/s00159-017-0106-5}

\bibitem[{{Hopkins} \& {Quataert}(2010)}]{2010MNRAS.405L..41H}
{Hopkins}, P.~F., \& {Quataert}, E. 2010, \mnras, 405, L41, \dodoi{10.1111/j.1745-3933.2010.00855.x}

\bibitem[{{Kacharov} {et~al.}(2018){Kacharov}, {Neumayer}, {Seth}, {Cappellari}, {McDermid}, {Walcher}, \& {B{\"o}ker}}]{2018MNRAS.480.1973K}
{Kacharov}, N., {Neumayer}, N., {Seth}, A.~C., {et~al.} 2018, \mnras, 480, 1973, \dodoi{10.1093/mnras/sty1985}

\bibitem[{{Kroupa}(2001)}]{2001MNRAS.322..231K}
{Kroupa}, P. 2001, \mnras, 322, 231, \dodoi{10.1046/j.1365-8711.2001.04022.x}

\bibitem[{{Kroupa}(2002)}]{2002Sci...295...82K}
---. 2002, Science, 295, 82, \dodoi{10.1126/science.1067524}

\bibitem[{{Lauer} {et~al.}(2012){Lauer}, {Bender}, {Kormendy}, {Rosenfield}, \& {Green}}]{2012ApJ...745..121L}
{Lauer}, T.~R., {Bender}, R., {Kormendy}, J., {Rosenfield}, P., \& {Green}, R.~F. 2012, \apj, 745, 121, \dodoi{10.1088/0004-637X/745/2/121}

\bibitem[{{Lauer} {et~al.}(1998){Lauer}, {Faber}, {Ajhar}, {Grillmair}, \& {Scowen}}]{1998AJ....116.2263L}
{Lauer}, T.~R., {Faber}, S.~M., {Ajhar}, E.~A., {Grillmair}, C.~J., \& {Scowen}, P.~A. 1998, \aj, 116, 2263, \dodoi{10.1086/300617}

\bibitem[{{Lauer} {et~al.}(1993){Lauer}, {Faber}, {Groth}, {Shaya}, {Campbell}, {Code}, {Currie}, {Baum}, {Ewald}, {Hester}, {Holtzman}, {Kristian}, {Light}, {Ligynds}, {O'Neil}, \& {Westphal}}]{1993AJ....106.1436L}
{Lauer}, T.~R., {Faber}, S.~M., {Groth}, E.~J., {et~al.} 1993, \aj, 106, 1436, \dodoi{10.1086/116737}

\bibitem[{{Li} {et~al.}(2011){Li}, {Garcia}, {Forman}, {Jones}, {Kraft}, {Lal}, {Murray}, \& {Wang}}]{2011ApJ...728L..10L}
{Li}, Z., {Garcia}, M.~R., {Forman}, W.~R., {et~al.} 2011, \apjl, 728, L10, \dodoi{10.1088/2041-8205/728/1/L10}

\bibitem[{{Li} {et~al.}(2023){Li}, {Li}, {Garc{\'\i}a-Benito}, \& {Jin}}]{2023ApJ...958...89L}
{Li}, Z., {Li}, Z., {Garc{\'\i}a-Benito}, R., \& {Jin}, Y. 2023, \apj, 958, 89, \dodoi{10.3847/1538-4357/ad0299}

\bibitem[{{Li} {et~al.}(2020){Li}, {Li}, {Smith}, \& {Gao}}]{2020ApJ...905..138L}
{Li}, Z., {Li}, Z., {Smith}, M. W.~L., \& {Gao}, Y. 2020, \apj, 905, 138, \dodoi{10.3847/1538-4357/abc5ba}

\bibitem[{{Li} {et~al.}(2019){Li}, {Li}, {Zhou}, {Gao}, {Jiang}, \& {Dong}}]{2019MNRAS.484..964L}
{Li}, Z., {Li}, Z., {Zhou}, P., {et~al.} 2019, \mnras, 484, 964, \dodoi{10.1093/mnras/stz040}

\bibitem[{{Li} {et~al.}(2009){Li}, {Wang}, \& {Wakker}}]{2009MNRAS.397..148L}
{Li}, Z., {Wang}, Q.~D., \& {Wakker}, B.~P. 2009, \mnras, 397, 148, \dodoi{10.1111/j.1365-2966.2009.14918.x}

\bibitem[{{Lockhart} {et~al.}(2018){Lockhart}, {Lu}, {Peiris}, {Rich}, {Bouchez}, \& {Ghez}}]{2018ApJ...854..121L}
{Lockhart}, K.~E., {Lu}, J.~R., {Peiris}, H.~V., {et~al.} 2018, \apj, 854, 121, \dodoi{10.3847/1538-4357/aaaa71}

\bibitem[{{MacLeod} {et~al.}(2012){MacLeod}, {Guillochon}, \& {Ramirez-Ruiz}}]{2012ApJ...757..134M}
{MacLeod}, M., {Guillochon}, J., \& {Ramirez-Ruiz}, E. 2012, \apj, 757, 134, \dodoi{10.1088/0004-637X/757/2/134}

\bibitem[{{Madigan} {et~al.}(2018){Madigan}, {Halle}, {Moody}, {McCourt}, {Nixon}, \& {Wernke}}]{2018ApJ...853..141M}
{Madigan}, A.-M., {Halle}, A., {Moody}, M., {et~al.} 2018, \apj, 853, 141, \dodoi{10.3847/1538-4357/aaa714}

\bibitem[{{Marigo} {et~al.}(2013){Marigo}, {Bressan}, {Nanni}, {Girardi}, \& {Pumo}}]{2013MNRAS.434..488M}
{Marigo}, P., {Bressan}, A., {Nanni}, A., {Girardi}, L., \& {Pumo}, M.~L. 2013, \mnras, 434, 488, \dodoi{10.1093/mnras/stt1034}

\bibitem[{{McDonald} {et~al.}(2011){McDonald}, {Johnson}, \& {Zijlstra}}]{2011MNRAS.416L...6M}
{McDonald}, I., {Johnson}, C.~I., \& {Zijlstra}, A.~A. 2011, \mnras, 416, L6, \dodoi{10.1111/j.1745-3933.2011.01086.x}

\bibitem[{{McDonald} \& {Zijlstra}(2015)}]{2015MNRAS.448..502M}
{McDonald}, I., \& {Zijlstra}, A.~A. 2015, \mnras, 448, 502, \dodoi{10.1093/mnras/stv007}

\bibitem[{{Melchior} \& {Combes}(2013)}]{2013A&A...549A..27M}
{Melchior}, A.~L., \& {Combes}, F. 2013, \aap, 549, A27, \dodoi{10.1051/0004-6361/201220204}

\bibitem[{{Melchior} \& {Combes}(2017)}]{2017A&A...607L...7M}
{Melchior}, A.-L., \& {Combes}, F. 2017, \aap, 607, L7, \dodoi{10.1051/0004-6361/201730910}

\bibitem[{{Melchior} \& {Combes}(2025)}]{2025A&A...693A..24M}
---. 2025, \aap, 693, A24, \dodoi{10.1051/0004-6361/202347597}

\bibitem[{{Menezes} {et~al.}(2013){Menezes}, {Steiner}, \& {Ricci}}]{2013ApJ...762L..29M}
{Menezes}, R.~B., {Steiner}, J.~E., \& {Ricci}, T.~V. 2013, \apjl, 762, L29, \dodoi{10.1088/2041-8205/762/2/L29}

\bibitem[{{Miglio} {et~al.}(2012){Miglio}, {Brogaard}, {Stello}, {Chaplin}, {D'Antona}, {Montalb{\'a}n}, {Basu}, {Bressan}, {Grundahl}, {Pinsonneault}, {Serenelli}, {Elsworth}, {Hekker}, {Kallinger}, {Mosser}, {Ventura}, {Bonanno}, {Noels}, {Silva Aguirre}, {Szabo}, {Li}, {McCauliff}, {Middour}, \& {Kjeldsen}}]{2012MNRAS.419.2077M}
{Miglio}, A., {Brogaard}, K., {Stello}, D., {et~al.} 2012, \mnras, 419, 2077, \dodoi{10.1111/j.1365-2966.2011.19859.x}

\bibitem[{{Mignone} {et~al.}(2007){Mignone}, {Bodo}, {Massaglia}, {Matsakos}, {Tesileanu}, {Zanni}, \& {Ferrari}}]{2007ApJS..170..228M}
{Mignone}, A., {Bodo}, G., {Massaglia}, S., {et~al.} 2007, \apjs, 170, 228, \dodoi{10.1086/513316}

\bibitem[{{Murchikova} {et~al.}(2019){Murchikova}, {Phinney}, {Pancoast}, \& {Blandford}}]{2019Natur.570...83M}
{Murchikova}, E.~M., {Phinney}, E.~S., {Pancoast}, A., \& {Blandford}, R.~D. 2019, \nat, 570, 83, \dodoi{10.1038/s41586-019-1242-z}

\bibitem[{{Neumayer} {et~al.}(2020){Neumayer}, {Seth}, \& {B{\"o}ker}}]{2020A&ARv..28....4N}
{Neumayer}, N., {Seth}, A., \& {B{\"o}ker}, T. 2020, \aapr, 28, 4, \dodoi{10.1007/s00159-020-00125-0}

\bibitem[{{Nguyen} {et~al.}(2022){Nguyen}, {Costa}, {Girardi}, {Volpato}, {Bressan}, {Chen}, {Marigo}, {Fu}, \& {Goudfrooij}}]{2022A&A...665A.126N}
{Nguyen}, C.~T., {Costa}, G., {Girardi}, L., {et~al.} 2022, \aap, 665, A126, \dodoi{10.1051/0004-6361/202244166}

\bibitem[{{Paumard} {et~al.}(2006){Paumard}, {Genzel}, {Martins}, {Nayakshin}, {Beloborodov}, {Levin}, {Trippe}, {Eisenhauer}, {Ott}, {Gillessen}, {Abuter}, {Cuadra}, {Alexander}, \& {Sternberg}}]{2006ApJ...643.1011P}
{Paumard}, T., {Genzel}, R., {Martins}, F., {et~al.} 2006, \apj, 643, 1011, \dodoi{10.1086/503273}

\bibitem[{{Peiris} \& {Tremaine}(2003)}]{2003ApJ...599..237P}
{Peiris}, H.~V., \& {Tremaine}, S. 2003, \apj, 599, 237, \dodoi{10.1086/378638}

\bibitem[{{Peng} {et~al.}(2023){Peng}, {Li}, {Sjouwerman}, {Yang}, {Jiang}, \& {Shen}}]{2023ApJ...953...12P}
{Peng}, S., {Li}, Z., {Sjouwerman}, L.~O., {et~al.} 2023, \apj, 953, 12, \dodoi{10.3847/1538-4357/acdddd}

\bibitem[{{Ploeckinger} \& {Schaye}(2020)}]{2020MNRAS.497.4857P}
{Ploeckinger}, S., \& {Schaye}, J. 2020, \mnras, 497, 4857, \dodoi{10.1093/mnras/staa2172}

\bibitem[{{Quataert}(2004)}]{2004ApJ...613..322Q}
{Quataert}, E. 2004, \apj, 613, 322, \dodoi{10.1086/422973}

\bibitem[{{Rau} {et~al.}(2018){Rau}, {Nielsen}, {Carpenter}, \& {Airapetian}}]{2018ApJ...869....1R}
{Rau}, G., {Nielsen}, K.~E., {Carpenter}, K.~G., \& {Airapetian}, V. 2018, \apj, 869, 1, \dodoi{10.3847/1538-4357/aaf0a0}

\bibitem[{{Ressler} {et~al.}(2018){Ressler}, {Quataert}, \& {Stone}}]{2018MNRAS.478.3544R}
{Ressler}, S.~M., {Quataert}, E., \& {Stone}, J.~M. 2018, \mnras, 478, 3544, \dodoi{10.1093/mnras/sty1146}

\bibitem[{{Ressler} {et~al.}(2020{\natexlab{a}}){Ressler}, {Quataert}, \& {Stone}}]{2020MNRAS.492.3272R}
---. 2020{\natexlab{a}}, \mnras, 492, 3272, \dodoi{10.1093/mnras/stz3605}

\bibitem[{{Ressler} {et~al.}(2020{\natexlab{b}}){Ressler}, {White}, {Quataert}, \& {Stone}}]{2020ApJ...896L...6R}
{Ressler}, S.~M., {White}, C.~J., {Quataert}, E., \& {Stone}, J.~M. 2020{\natexlab{b}}, \apjl, 896, L6, \dodoi{10.3847/2041-8213/ab9532}

\bibitem[{{Rose} {et~al.}(2023){Rose}, {Naoz}, {Sari}, \& {Linial}}]{2023ApJ...955...30R}
{Rose}, S.~C., {Naoz}, S., {Sari}, R., \& {Linial}, I. 2023, \apj, 955, 30, \dodoi{10.3847/1538-4357/acee75}

\bibitem[{{Rosenfield} {et~al.}(2012){Rosenfield}, {Johnson}, {Girardi}, {Dalcanton}, {Bressan}, {Lang}, {Williams}, {Guhathakurta}, {Howley}, {Lauer}, {Bell}, {Bianchi}, {Caldwell}, {Dolphin}, {Dorman}, {Gilbert}, {Kalirai}, {Larsen}, {Olsen}, {Rix}, {Seth}, {Skillman}, \& {Weisz}}]{2012ApJ...755..131R}
{Rosenfield}, P., {Johnson}, L.~C., {Girardi}, L., {et~al.} 2012, \apj, 755, 131, \dodoi{10.1088/0004-637X/755/2/131}

\bibitem[{{Russell} {et~al.}(2017){Russell}, {Wang}, \& {Cuadra}}]{2017MNRAS.464.4958R}
{Russell}, C. M.~P., {Wang}, Q.~D., \& {Cuadra}, J. 2017, \mnras, 464, 4958, \dodoi{10.1093/mnras/stw2584}

\bibitem[{{Saglia} {et~al.}(2010){Saglia}, {Fabricius}, {Bender}, {Montalto}, {Lee}, {Riffeser}, {Seitz}, {Morganti}, {Gerhard}, \& {Hopp}}]{2010A&A...509A..61S}
{Saglia}, R.~P., {Fabricius}, M., {Bender}, R., {et~al.} 2010, \aap, 509, A61, \dodoi{10.1051/0004-6361/200912805}

\bibitem[{{Shcherbakov} {et~al.}(2014){Shcherbakov}, {Wong}, {Irwin}, \& {Reynolds}}]{2014ApJ...782..103S}
{Shcherbakov}, R.~V., {Wong}, K.-W., {Irwin}, J.~A., \& {Reynolds}, C.~S. 2014, \apj, 782, 103, \dodoi{10.1088/0004-637X/782/2/103}

\bibitem[{{Shi} {et~al.}(2021){Shi}, {Li}, {Yuan}, \& {Zhu}}]{2021NatAs...5..928S}
{Shi}, F., {Li}, Z., {Yuan}, F., \& {Zhu}, B. 2021, Nature Astronomy, 5, 928, \dodoi{10.1038/s41550-021-01394-0}

\bibitem[{{Shi} {et~al.}(2022){Shi}, {Zhu}, {Li}, \& {Yuan}}]{2022ApJ...926..209S}
{Shi}, F., {Zhu}, B., {Li}, Z., \& {Yuan}, F. 2022, \apj, 926, 209, \dodoi{10.3847/1538-4357/ac4789}

\bibitem[{{Su} {et~al.}(2022){Su}, {Li}, {Hou}, {Zhang}, \& {Cheng}}]{2022MNRAS.516.1788S}
{Su}, Z., {Li}, Z., {Hou}, M., {Zhang}, M., \& {Cheng}, Z. 2022, \mnras, 516, 1788, \dodoi{10.1093/mnras/stac2345}

\bibitem[{{Sullivan} {et~al.}(2006){Sullivan}, {Le Borgne}, {Pritchet}, {Hodsman}, {Neill}, {Howell}, {Carlberg}, {Astier}, {Aubourg}, {Balam}, {Basa}, {Conley}, {Fabbro}, {Fouchez}, {Guy}, {Hook}, {Pain}, {Palanque-Delabrouille}, {Perrett}, {Regnault}, {Rich}, {Taillet}, {Baumont}, {Bronder}, {Ellis}, {Filiol}, {Lusset}, {Perlmutter}, {Ripoche}, \& {Tao}}]{2006ApJ...648..868S}
{Sullivan}, M., {Le Borgne}, D., {Pritchet}, C.~J., {et~al.} 2006, \apj, 648, 868, \dodoi{10.1086/506137}

\bibitem[{{Tailo} {et~al.}(2021){Tailo}, {Milone}, {Lagioia}, {D'Antona}, {Jang}, {Vesperini}, {Marino}, {Ventura}, {Caloi}, {Carlos}, {Cordoni}, {Dondoglio}, {Mohandasan}, {Nastasio}, \& {Legnardi}}]{2021MNRAS.503..694T}
{Tailo}, M., {Milone}, A.~P., {Lagioia}, E.~P., {et~al.} 2021, \mnras, 503, 694, \dodoi{10.1093/mnras/stab568}

\bibitem[{{Tremaine}(1995)}]{1995AJ....110..628T}
{Tremaine}, S. 1995, \aj, 110, 628, \dodoi{10.1086/117548}

\bibitem[{{Vink} {et~al.}(2001){Vink}, {de Koter}, \& {Lamers}}]{2001A&A...369..574V}
{Vink}, J.~S., {de Koter}, A., \& {Lamers}, H.~J.~G.~L.~M. 2001, \aap, 369, 574, \dodoi{10.1051/0004-6361:20010127}

\bibitem[{{Walcher} {et~al.}(2006){Walcher}, {B{\"o}ker}, {Charlot}, {Ho}, {Rix}, {Rossa}, {Shields}, \& {van der Marel}}]{2006ApJ...649..692W}
{Walcher}, C.~J., {B{\"o}ker}, T., {Charlot}, S., {et~al.} 2006, \apj, 649, 692, \dodoi{10.1086/505166}

\bibitem[{{Wang} {et~al.}(2013){Wang}, {Nowak}, {Markoff}, {Baganoff}, {Nayakshin}, {Yuan}, {Cuadra}, {Davis}, {Dexter}, {Fabian}, {Grosso}, {Haggard}, {Houck}, {Ji}, {Li}, {Neilsen}, {Porquet}, {Ripple}, \& {Shcherbakov}}]{2013Sci...341..981W}
{Wang}, Q.~D., {Nowak}, M.~A., {Markoff}, S.~B., {et~al.} 2013, Science, 341, 981, \dodoi{10.1126/science.1240755}

\bibitem[{{Weinberger} {et~al.}(2017){Weinberger}, {Springel}, {Hernquist}, {Pillepich}, {Marinacci}, {Pakmor}, {Nelson}, {Genel}, {Vogelsberger}, {Naiman}, \& {Torrey}}]{2017MNRAS.465.3291W}
{Weinberger}, R., {Springel}, V., {Hernquist}, L., {et~al.} 2017, \mnras, 465, 3291, \dodoi{10.1093/mnras/stw2944}

\bibitem[{{Weinberger} {et~al.}(2018){Weinberger}, {Springel}, {Pakmor}, {Nelson}, {Genel}, {Pillepich}, {Vogelsberger}, {Marinacci}, {Naiman}, {Torrey}, \& {Hernquist}}]{2018MNRAS.479.4056W}
{Weinberger}, R., {Springel}, V., {Pakmor}, R., {et~al.} 2018, \mnras, 479, 4056, \dodoi{10.1093/mnras/sty1733}

\bibitem[{{Wernke} \& {Madigan}(2019)}]{2019ApJ...880...42W}
{Wernke}, H.~N., \& {Madigan}, A.-M. 2019, \apj, 880, 42, \dodoi{10.3847/1538-4357/ab2711}

\bibitem[{{Wood} {et~al.}(2024){Wood}, {Harper}, \& {M{\"u}ller}}]{2024ApJ...967..120W}
{Wood}, B.~E., {Harper}, G.~M., \& {M{\"u}ller}, H.-R. 2024, \apj, 967, 120, \dodoi{10.3847/1538-4357/ad401f}

\bibitem[{{Yang} {et~al.}(2017){Yang}, {Li}, {Sjouwerman}, {Yuan}, \& {Shen}}]{2017ApJ...845..140Y}
{Yang}, Y., {Li}, Z., {Sjouwerman}, L.~O., {Yuan}, F., \& {Shen}, Z.-Q. 2017, \apj, 845, 140, \dodoi{10.3847/1538-4357/aa8265}

\bibitem[{{Yao} \& {Gan}(2020)}]{2020MNRAS.492..444Y}
{Yao}, Z., \& {Gan}, Z. 2020, \mnras, 492, 444, \dodoi{10.1093/mnras/stz3474}

\bibitem[{{Yoon} {et~al.}(2019){Yoon}, {Yuan}, {Ostriker}, {Ciotti}, \& {Zhu}}]{2019ApJ...885...16Y}
{Yoon}, D., {Yuan}, F., {Ostriker}, J.~P., {Ciotti}, L., \& {Zhu}, B. 2019, \apj, 885, 16, \dodoi{10.3847/1538-4357/ab45e8}

\bibitem[{{Yuan} {et~al.}(2015){Yuan}, {Gan}, {Narayan}, {Sadowski}, {Bu}, \& {Bai}}]{2015ApJ...804..101Y}
{Yuan}, F., {Gan}, Z., {Narayan}, R., {et~al.} 2015, \apj, 804, 101, \dodoi{10.1088/0004-637X/804/2/101}

\bibitem[{{Yuan} \& {Narayan}(2014)}]{2014ARA&A..52..529Y}
{Yuan}, F., \& {Narayan}, R. 2014, \araa, 52, 529, \dodoi{10.1146/annurev-astro-082812-141003}

\bibitem[{{Zhang} {et~al.}(2019){Zhang}, {Wang}, {Foster}, {Sun}, {Li}, \& {Ji}}]{2019ApJ...885..157Z}
{Zhang}, S., {Wang}, Q.~D., {Foster}, A.~R., {et~al.} 2019, \apj, 885, 157, \dodoi{10.3847/1538-4357/ab4a0f}

\end{thebibliography}
\bibliographystyle{aasjournal}



\end{document}